\documentclass[amsmath,amssymb,prd,twocolumn,final,superscriptaddress]{revtex4-2}
\pdfoutput=1
\usepackage[english]{babel}
\usepackage[utf8]{inputenc}
\usepackage[T1]{fontenc}
\usepackage[colorlinks = true,
            linkcolor = blue,
            urlcolor  = blue,
            citecolor = blue,
            anchorcolor = blue]{hyperref}
\usepackage{graphicx}
\usepackage{autonum}
\usepackage{orcidlink}
\usepackage{placeins}

\newcommand{\lrfrac}[2]{\left( \frac{#1}{#2}\right)}

\begin{document}
\title{
Chaotic dynamics and resonances in Einstein-Rosen standing wave spacetimes
}

\author{Adam A. Zychowicz\,\orcidlink{0009-0004-2544-0632}}
\affiliation{Astronomical Observatory, Jagiellonian University}
\affiliation{Doctoral School of Exact and Natural Sciences, Jagiellonian University}

\begin{abstract}

Recent work has revealed chaotic geodesic behavior in Einstein-Rosen standing gravitational wave spacetimes. The dynamics of massive test particles are dominated by chaotic heteroclinic networks. This study demonstrates that the dynamical structure of the system is far richer than previously thought. Applying the subharmonic Melnikov function method to Poincaré maps reveals the admissible types of heteroclinic networks. I trace the evolution of these structures under the continuous change of the solution's parameters. In this way, I identify bifurcation phenomena, including a period-doubling bifurcation.

\end{abstract}
\maketitle{}

\section{Introduction}
Standing waves are of fundamental importance across many fields of physics. They are essential in a wide range of physical phenomena and have numerous practical engineering applications in daily life. In the fields described by linear theories, standing waves can usually be easily constructed as a superposition of two simple traveling waves. In the general theory of relativity, the linearized regime is relatively well understood, but the nonlinearity of the full Einstein equations makes it an interesting but difficult subject to study. One way to make it possible to work with the full nonlinear Einstein equations in the context of standing gravitational waves is to focus on solutions with cylindrical symmetry, which simplify the equations considerably.

Over the past few years, standing gravitational waves (SGWs) have been the subject of numerous articles exploring their properties in various contexts, including the fundamental question of the definition of SGWs \cite{Stephani2003definition, SzybkaCieslik2019definition}, properties of cosmological models of SGWs \cite{SzybkaNaqvi2021freefall,GlodSikoraSzybka2022cosmological}, and differences between known solutions in the linear weak-field regime \cite{NikielSzybka2025linear}. In \cite{SzybkaNaqvi2023chaos}, the authors show evidence of chaotic behavior of numerically obtained geodesics in an example cylindrically symmetric Einstein-Rosen SGW spacetime using Poincar\'e sections. In recent work \cite{backreaction}, the high-frequency limit of the Halilsoy \cite{Halilsoy:1988} and Chandrasekhar \cite{Chandrasekhar1986} SGW solutions was examined in the context of backreaction within the Green-Wald framework \cite{greenwald}.

In this paper I extend the work of \cite{SzybkaNaqvi2023chaos}. I~examine geodesics of massive test particles in a family of solutions belonging to the Einstein-Rosen SGW class. I~demonstrate that the previously identified features are only a part of a far richer structure. Using the subharmonic Melnikov method, I identify the resonant chains of the constructed Poincar\'e maps. In \cite{backreaction}, a one-parameter family of solutions fulfilling the requirements of the Green-Wald framework was constructed, and its backreaction effect was computed. The Einstein-Rosen SGW is included as a special case of the Halilsoy solution, so a high-frequency limit similar to the one presented in \cite{backreaction} for the Halilsoy solution can also be constructed for the Einstein-Rosen SGW. The effective spacetime arising as this high-frequency limit does not exhibit chaotic behavior. This raises the question of how chaotic behavior changes in this limit. To investigate this, I trace how a continuous change of parameters along the family of solutions constructed in \cite{backreaction} affects the Poincar\'e sections, and I identify bifurcation phenomena, including a period-doubling bifurcation.
\section{Setup} \label{sec:setup}
I consider a cylindrically symmetric metric of the form: 
\begin{equation}
\label{eq:metricER}
    ds^2=e^{2(\gamma-\psi)}(-dt^2+d\rho^2)+\rho^2e^{-2\psi}d\phi^2+e^{2\psi}dz^2,
\end{equation}
with $\rho>0;\; t,z \in (-\infty,\infty);\; \phi\in [0,2\pi)$. The functions $\gamma=\gamma(t,\rho)$ and $\psi=\psi(t,\rho)$ are functions of the $t$ and $\rho$ coordinates only. The metric (\ref{eq:metricER}) is not the most general form of a cylindrically symmetric metric, but is the one that will be used throughout this work. The vacuum Einstein field equations (EFEs) of the metric (\ref{eq:metricER}) reduce to three equations. The first equation requires $\psi(t,\rho)$ to satisfy the wave equation in cylindrical coordinates:
\begin{equation}
\psi''+\frac{1}{\rho}\psi'-\ddot{\psi}=0,    
\end{equation}
where the prime ( $'$ ) denotes $\partial_\rho$ and the dot ( $\dot{}$ ) denotes $\partial_t$. The remaining EFEs give $\gamma$ from $\psi$ by quadratures:
\begin{equation}
\dot{\gamma}=2 \rho \dot{\psi}\psi'     ,
\end{equation}
\begin{equation}
\gamma'=\rho(\dot{\psi}^2+\psi'^2)      .
\end{equation}

The function $\gamma$ is closely related to Kip Thorne's C-energy \cite{Thorne1965} (the C-energy is equal to $\gamma/4$). This family of metrics contains the class of Einstein-Rosen gravitational waves \cite{ER1937}. I consider a specific solution of the vacuum EFEs studied in \cite{SzybkaNaqvi2023chaos}, belonging to the Einstein-Rosen class, in the form of a cylindrical standing gravitational wave \cite{Stephani2003definition}:
\begin{gather}\label{eq:ER_SGW_psigamma}
\begin{split}
&\qquad\qquad\qquad\quad\psi(t,\rho)=A \frac{1}{2}J_0\lrfrac{\rho}{\lambda} \cos{\lrfrac{t}{\lambda}},\\
&\gamma(t,\rho)=\\&A^2 \frac{\rho^2}{8 \lambda^2} \Biggl[J_0^2\lrfrac{\rho}{\lambda}+J_1^2\lrfrac{\rho}{\lambda} -2 \frac{\lambda}{\rho} J_0\lrfrac{\rho}{\lambda} J_1\lrfrac{\rho}{\lambda} \cos^2{\lrfrac{t}{\lambda}}  \Biggr],
\end{split}
\end{gather}
where $J_n$ are Bessel functions of the first kind of order~$n$. On the axis $\rho=0$, the cylindrical coordinate system is not well defined, but the spacetime remains regular. Let $\tau$ be a proper time parameter along a timelike geodesic. Then $u^\alpha=\frac{dx^\alpha}{d\tau}$ are the components of the four-velocity of a massive test particle ($u^\alpha u_\alpha=-1$). The Killing vector fields $\xi=\partial_z$ and $\eta=\partial_\phi$ give rise to two conserved quantities along geodesics:
\begin{equation}
u_z=g_{\alpha\beta}u^\alpha\xi^{\beta}, \quad \quad l=g_{\alpha\beta}u^\alpha\eta^{\beta}=\rho^2e^{-2\psi} u^\phi.
\end{equation}
Throughout this paper only trajectories with $u_z=0$, that is, $z=\text{const}$, will be examined. The conserved quantity $l$ can be understood as angular momentum per unit mass. The geodesic equation then can be reduced to a first order system of four equations describing the evolution of the coordinates and their derivatives along the geodesic: $t(\tau)$, $T(\tau)\equiv\dot{t}$, $\rho(\tau)$, and $P(\tau)\equiv \dot{\rho}$, which then can be solved numerically (see for example \cite{SzybkaNaqvi2023chaos}):
\begin{gather}\label{eq:geodesicEQs} 
\begin{split}
\dot{t}&=T,\\
\dot{T}&=-(T^2+P^2)\Delta_{,t}-2TP\Delta_{,\rho}+e^{-2\Delta+2\psi}\psi_{,t}\frac{l^2}{\rho^2},\\
\dot{\rho}&=P,\\
\dot{P}&=-(T^2+P^2)\Delta_{,\rho}-2TP\Delta_{,t}-e^{-2\Delta+2\psi}\left(\psi_{,\rho}-\frac{1}{\rho}\right)\frac{l^2}{\rho^2},\\
\end{split}
\end{gather}
where $\Delta=\gamma-\psi$ and the dot denotes the derivative with respect to proper time $\tau$ along the geodesic. The evolution of $\phi(\tau)$ separates, so $\phi(\tau)$ can be found by quadratures after the rest of the system is solved using $\dot{\phi}=u^\phi=l\rho^{-2}e^{2\psi}$. For non-trivial parameters, no closed-form solution for the system (\ref{eq:geodesicEQs}) is known, so geodesics will be computed numerically. The numerical computations were performed using Wolfram Mathematica \mbox{NDSolve} procedure with the default choice of method and solver parameters \cite{Mathematica}. One way to examine the behavior of the geodesics defined by equations (\ref{eq:geodesicEQs}) is to represent it as a Poincar\'e map and study the Poincar\'e section $t\,\text{mod}\,2\pi \lambda=0$. The following section explains how to construct the Poincaré map and obtain the corresponding Poincaré section.

\section{Construction of Poincar\'e maps} \label{sec:poincare}
While not strictly necessary for numerical integration, it is worth noting that the normalization $u^\alpha u_\alpha=-1$ can be used as another first integral eliminating one of the four dynamical variables (for example $T$). Taking this constraint into account, the resulting dynamical system can be thought of as being three-dimensional ($t,\rho,P$). The system is chaotic, exhibiting sensitivity to initial conditions, as demonstrated in \cite{SzybkaNaqvi2023chaos}. It will be examined using the Poincar\'e sections. The wave described by the metric (\ref{eq:metricER}) and the solution (\ref{eq:ER_SGW_psigamma}) oscillates periodically with a period of $2\pi \lambda$. Due to the periodicity of the system, the $t$ coordinate itself can be treated as a periodic coordinate with this period. The behavior of the system can then be studied by examining the two dimensional plane of $(\rho,P)$ on the section $t\,\text{mod}\,2\pi\lambda=0$. Starting from some initial condition, this section allows one to track the coordinate distance from the symmetry axis and the radial velocity of the test particles after each period of the wave. This function mapping points from the two dimensional subspace of phase space $(\rho,P)$ onto itself defines the Poincar\'e map $\phi$ (also called a first recurrence map, or, in the case of a map defined by a specific phase of the driving force, like in this case, a stroboscopic map). The points reached by a test particle after one period can be effectively treated as a new initial condition. Integrating the trajectories longer than one period of the wave allows one to create multiple compositions of the resulting Poincar\'e map with itself, providing a tool for visualization of some of the behavior of the system. This composition of the map $\phi$ with itself $n$ times will be denoted $\phi^n$. The next section includes discussions of structures revealed using the map $\phi$ and the resulting Poincar\'e section.

\section{Layered heteroclinic networks}
\label{sec:layered}
\begin{figure}[]
\centering
\includegraphics[width=0.5\textwidth]{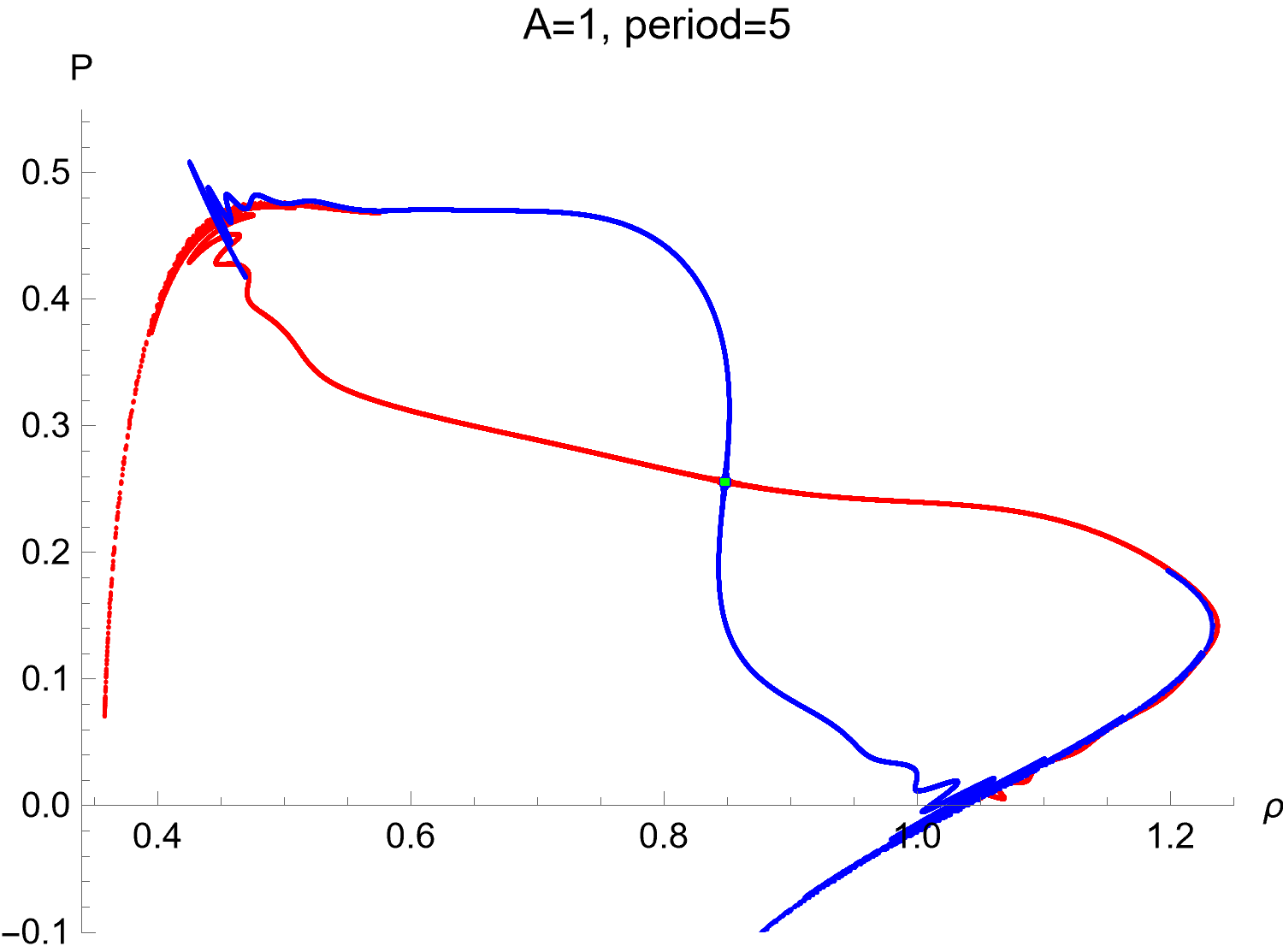}
\caption{Stable manifold (red) and unstable manifold (blue) of a hyperbolic fixed point of the map $\phi^5$ in the vicinity of the set of initial conditions marked in green.}
\label{fig:manifolds}
\end{figure}

In \cite{SzybkaNaqvi2023chaos}, a chaotic heteroclinic network consisting of five 5-periodic hyperbolic and corresponding five 5-periodic elliptic points was identified in the map $\phi$, which are hyperbolic and elliptic fixed points of $\phi^5$, respectively, with $\phi$ defined as in the previous section. I begin with an independent reproduction of the results of \cite{SzybkaNaqvi2023chaos}. To facilitate the comparison, in this section, the values $A=1$, $\lambda=\frac{1}{2\pi}$ and $l=1/3$ are used in the numerical integration of the geodesic equation. The unstable and stable manifolds of each hyperbolic fixed point can be visualized by evolving a set of initial conditions in the neighborhood of a given fixed point both forward and backward in time. Figure~\ref{fig:manifolds} shows a numerically obtained picture of stable and unstable manifolds of an example hyperbolic fixed point of $\phi^5$ belonging to the network discussed in \cite{SzybkaNaqvi2023chaos}. Stable and unstable manifolds cross in the vicinity of neighboring hyperbolic fixed points, creating a heteroclinic tangle, strong evidence of chaotic behavior. Every fixed point is tangled in the same way with two neighboring fixed points, and trajectories starting from the vicinity of the (un)stable manifold of any fixed point chaotically explore the entire 5-point network.

\begin{figure}[]
\centering
\includegraphics[width=0.4\textwidth]{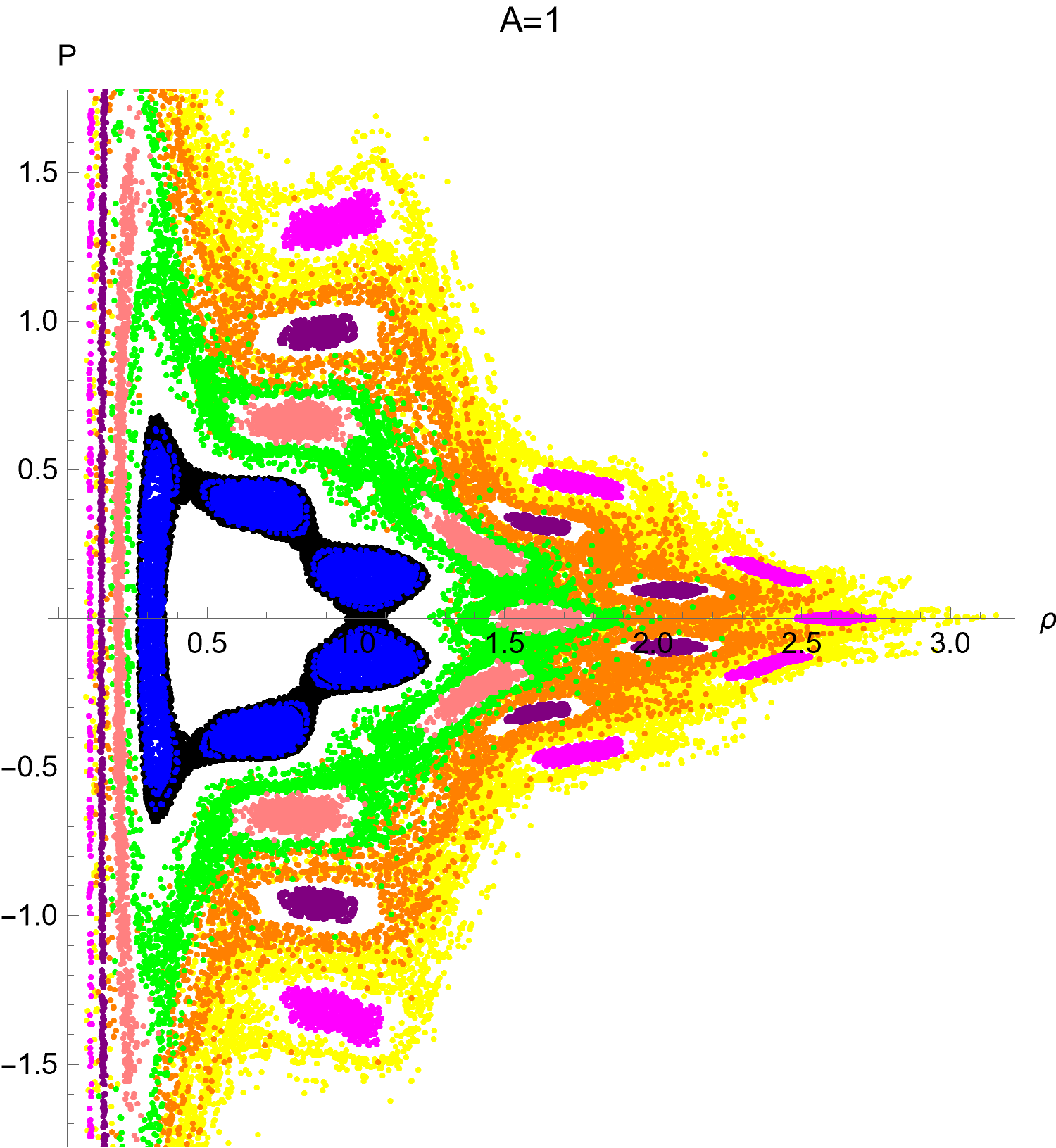}
\caption{The Poincar\'e sections of $t = 2\pi \lambda k$, for $k=0,1,2,...$, with parameters $A=1$, $\lambda=\frac{1}{2\pi}$ and $l=1/3$. The 5-periodic network identified in \cite{SzybkaNaqvi2023chaos} is visible and marked in black, with the corresponding set of neighborhoods of elliptic fixed points of $\phi^5$ marked in blue. Different colors mark sets of trajectories starting with similar conditions, either in the vicinity of a hyperbolic or elliptic fixed point of the $\phi^n$ map for different values of $n$.}
\label{fig:layered_external}
\end{figure}

I extend the results obtained in the article \cite{SzybkaNaqvi2023chaos} and show that the phase space is much more complicated, and many more networks of various periodicity can be identified. Figure \ref{fig:layered_external} shows a Poincar\'e map with many separate networks highlighted. The 5-periodic network discussed in \cite{SzybkaNaqvi2023chaos} is still clearly visible, but numerous additional structures can be identified. On the outside of the 5-periodic network, one can identify an entire family of networks of very similar behavior and structure, but of increasing periodicity. Subsequent networks are increasing in period by one with each step. The network adjacent to the 5-periodic network exhibits a period of 6 (green in Figure \ref{fig:layered_external}), the next one exhibits a period of 7 (orange), and so on. Between the hyperbolic fixed points, sets of elliptic fixed points exist, allowing for the existence of regions of quasi-periodic, non-chaotic motion. No sign of this pattern stopping was found with numerical examination, but identification of all the fixed points in the numerically obtained Poincar\'e map gets progressively harder with each ``layer'', especially for increasingly highly compacted regions around elliptic fixed points close to $\rho=0$, also visible in Figure \ref{fig:layered_external}.

\begin{figure}
\centering
\includegraphics[width=0.4\textwidth]{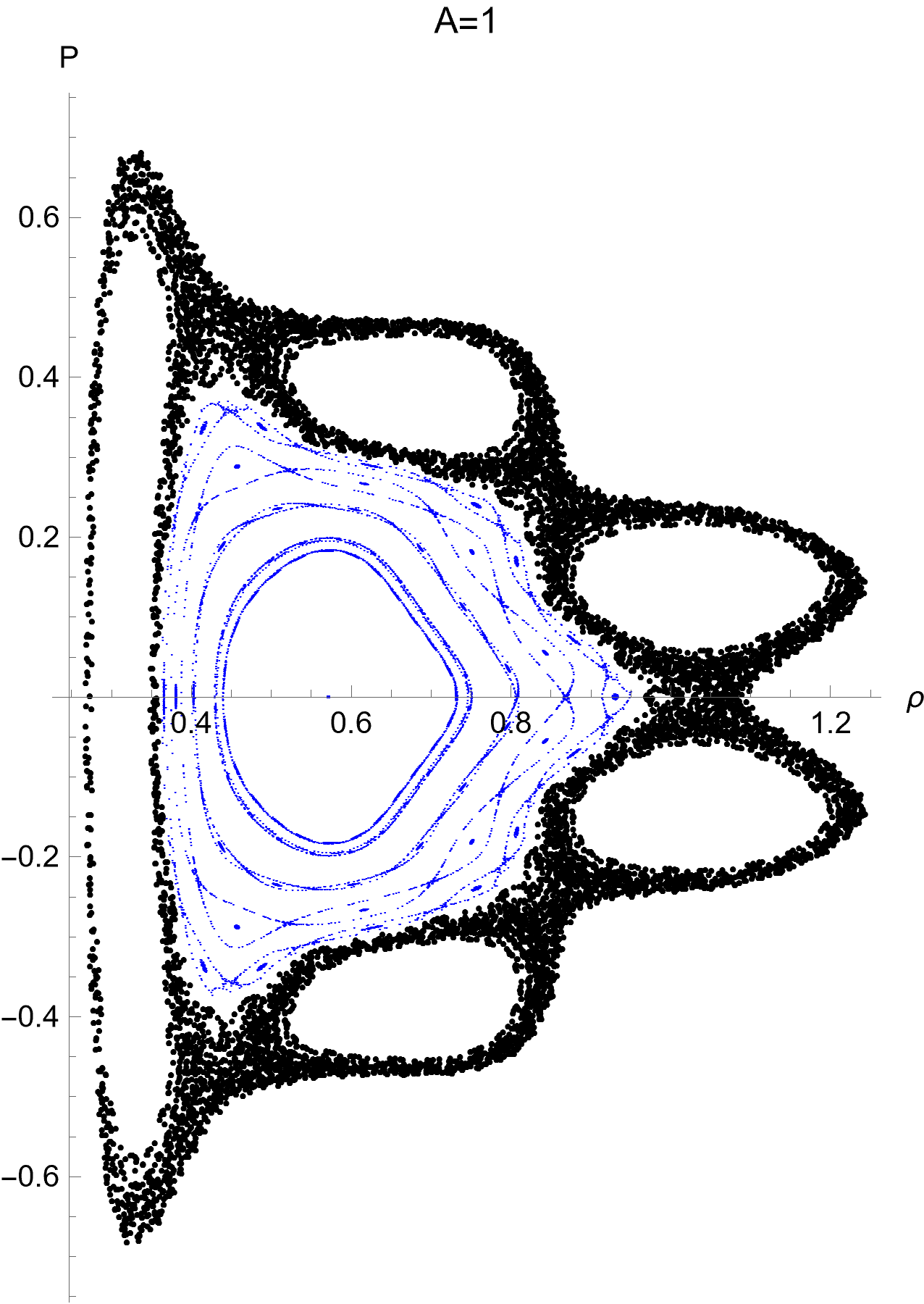}
\caption{The Poincar\'e sections of $t = 2\pi \lambda k$, for $k=0,1,2,...$, with parameters $A=1$, $\lambda=\frac{1}{2\pi}$ and $l=1/3$. The 5-periodic network identified in \cite{SzybkaNaqvi2023chaos} is shown in black for reference. Heteroclinic networks marked in blue exhibit different, non-sequential periodicities, with higher periods compared to external networks shown in Figure \ref{fig:layered_external}.}
\label{fig:layered_internal}
\end{figure}

In the center of the structure, there exists an elliptic fixed point $p_c\approx(0.5714,0)$ of the map $\phi$, with a subset of initial conditions following quasi-periodic motion in the vicinity of this point. In that case, it is natural to expect to also see the networks of lower periodicity inside the reference 5-periodic network, like period 4, 3, or 2. Surprisingly, no such thing is observed. Instead, closer examination reveals a complicated pattern of chaotic heteroclinic networks. Figure \ref{fig:layered_internal} depicts some of them, but many, possibly infinitely many more similar networks exist. These networks exhibit different, non-sequential, and non-monotonic periodicities, with higher periods compared to external neighboring networks. For example, the outermost of the networks marked in blue in Figure \ref{fig:layered_internal} has a period of 14 followed by 9, and then 13. It's highly probable that there are many more networks in this region of phase space. This irregular behavior and chaotic nature of the system make it hard to systematically detect them purely by visual examination of numerically obtained Poincar\'e section. The subharmonic Melnikov function method described in the following section will be used instead.

\section{Subharmonic Melnikov function analysis}

\begin{figure}[]
\centering
\includegraphics[width=0.49\textwidth]{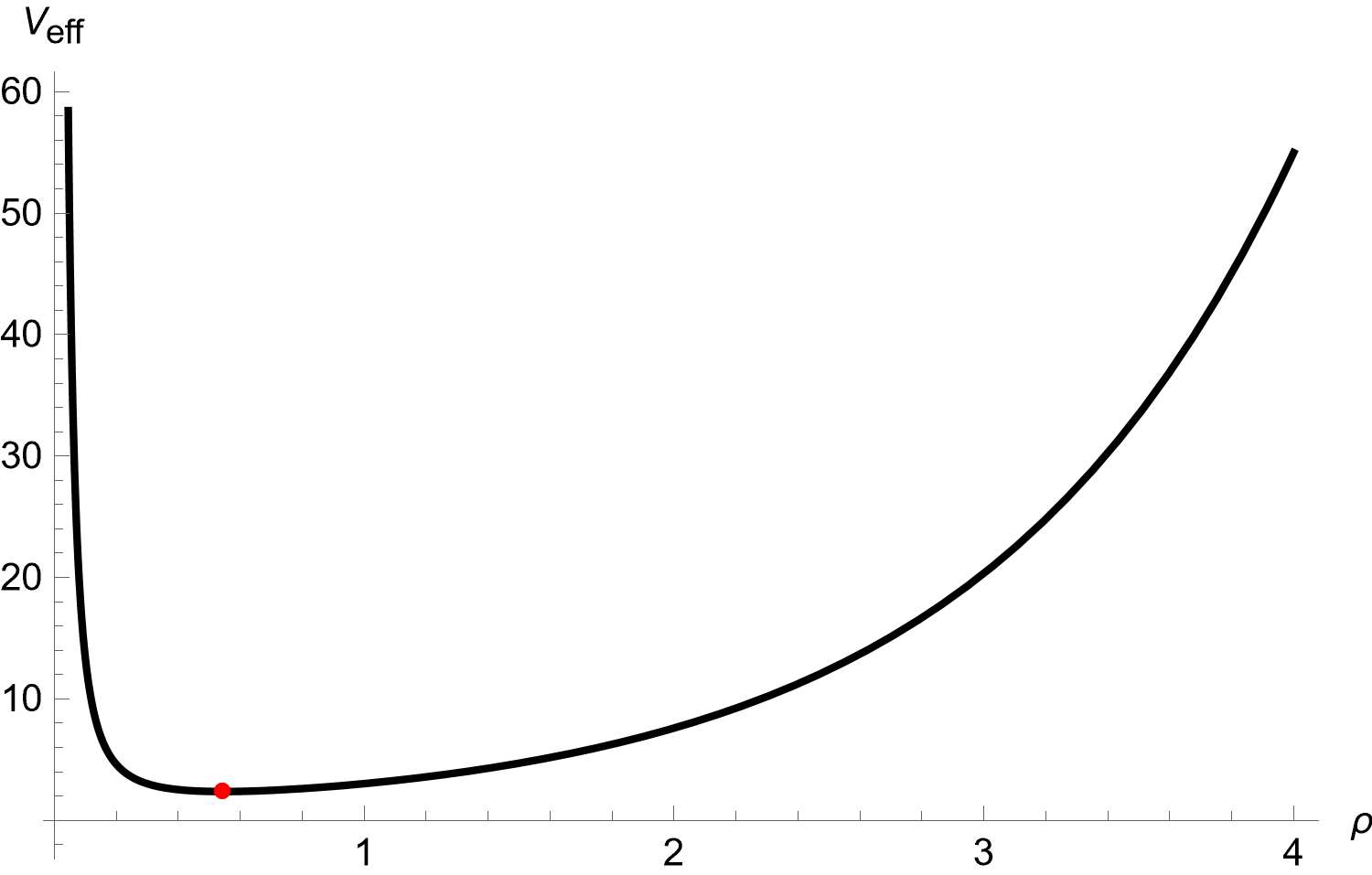}

\caption{Effective potential $V_{\text{eff}}$ included in the exactly integrable background Hamiltonian $H_0$. The red dot marks the minimum position at $\rho\approx0.545$ for $A=1$, $\lambda=\frac{1}{2\pi}$, $l=1/3$ and $\beta^2=\frac{\pi}{2}$.}
\label{fig:Veff}
\end{figure}

\begin{figure}[]
\centering
\includegraphics[width=0.49\textwidth]{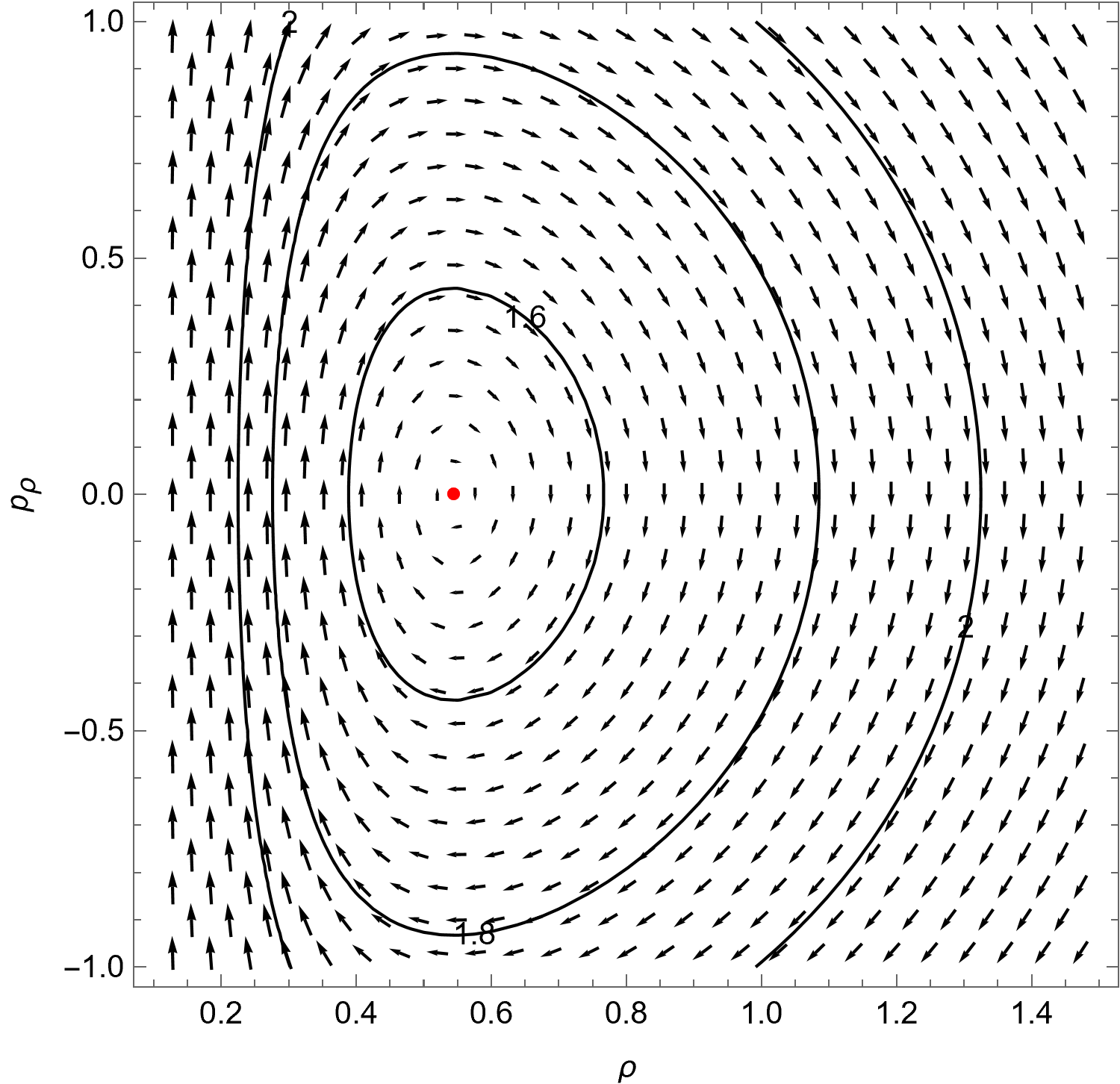}

\caption{Phase-space portrait of the system described by eq.\ (\ref{eq:H0eom}). The vector field shows the Hamiltonian flow. The contours show the level curves of constant energy $H_0=E$ of different example values (1.6, 1.8, 2). The red dot is a fixed point corresponding to the minimum of $V_{\text{eff}}$.}
\label{fig:phasespace}
\end{figure}

The sets of periodic points were found by manual examination of the obtained Poincar\'e maps. In this section, the structure of the networks they create is explained with the help of the so-called subharmonic Melnikov method \cite{Guckenheimer1983}. To use this method, the geodesic equation first needs to be represented using the Hamiltonian formulation. The conserved quantities are still $u_z=0$ and $l=\rho^2 e^{-2\psi}u^\phi$. The normalization of the four velocity $u^\alpha u_\alpha=-1$ gives:
\begin{equation}
    e^{2\psi}\frac{l^2}{\rho^2}-e^{2\Delta} (u^t)^2+e^{2\Delta}(u^\rho)^2=-1.
\end{equation}
This constraint can be rewritten using the canonical momentum $p_\alpha=g_{\alpha\beta}u^\beta$. Solving for $p_t$ gives rise to the reduced Hamiltonian:
\begin{equation}
    H=-p_t=\sqrt{p_\rho^2+e^{2\Delta}\left(1+e^{2\psi}\frac{l^2}{\rho^2}\right)}.
\end{equation}
To properly use the Melnikov method, the Hamiltonian needs to be split into two parts: $H_0$ - the unperturbed, integrable background, and $H_1$ - the leftover perturbation part periodic in $t$, so that $H(\rho,p_\rho,t)=H_0(\rho,p_\rho)+H_1(\rho,p_\rho,t)$. Motivated by the recent work of \cite{backreaction}, the unperturbed background was chosen to be the effective spacetime constructed as the high-frequency limit, construction of which is described in more detail in section~\ref{sec:HFL}. The background Hamiltonian $H_0$ is: 
\begin{align}
    H_0&=\sqrt{p_\rho^2+V_{\text{eff}}(\rho)},\\
    V_{\text{eff}}(\rho)&=e^{2\beta^2 \rho/\pi}\left(1+\frac{l^2}{\rho^2}\right).
\end{align}
The perturbation part is simply defined as \mbox{$H_1(\rho,p_\rho,t)=H(\rho,p_\rho,t)-H_0(\rho,p_\rho)$}. The potential $V_{\text{eff}}(\rho)$ has one extremum (minimum), and its plot is presented in Figure \ref{fig:Veff}. For the values of the parameters $A=1$, $\lambda=\frac{1}{2\pi}$, $l=1/3$ and $\beta^2=\frac{\pi}{2}$, the one minimum in the effective potential is located at $\rho_c\approx0.545$. 
The equations of motion of $H_0$ are:
\begin{equation}
\label{eq:H0eom}
\begin{aligned}
    \frac{d\rho}{dt}&=\frac{p_{\rho }}{\sqrt{e^{\frac{2 \beta ^2 \rho }{\pi }} \left(\frac{l^2}{\rho ^2}+1\right)+p_{\rho }{}^2}},\\
   \frac{dp_\rho}{dt}&=-\frac{e^{\frac{2 \beta ^2 \rho }{\pi }}\frac{\beta^2}{\pi} \left(1+\frac{l^2 }{\rho^2} -\frac{l^2\pi}{\rho^3\beta^2}\right)}{\sqrt{e^{\frac{2 \beta ^2 \rho }{\pi }} \left(\frac{l^2}{\rho
   ^2}+1\right)+p_{\rho }{}^2}}.
\end{aligned}    
\end{equation}
The dynamics of this system are visualized as a phase-space portrait in Figure \ref{fig:phasespace}. The system has only one fixed point, corresponding to the minimum of the potential $V_{\text{eff}}$.

The Melnikov method describes Hamiltonian systems of the form $H=H_0+\epsilon H_1$. For the unperturbed system described by $H_0$, for each relevant energy level $E=H_0(\rho,p_p)$, there exists a periodic orbit of period $T$. For each of those orbits, its period can be found, creating a relation $T(E)$. The full, perturbed system described by $H$ is being driven periodically with period $T_d$ (for the parameters used so far, $T_d=1$). One can impose the resonance condition for any co-prime pair $(m,n)$:
\begin{equation}
\label{eq:resonance}
    m T(E)=nT_d \Leftrightarrow T(E)=\frac{n}{m}T_d.
\end{equation}
For each such pair $(m,n)$ satisfying the resonance condition, there exists a specific value of the energy $E$. Let $q^{E}(t)$ be a periodic orbit of the unperturbed system described with the Hamiltonian $H_0$, with period $T(E)$. The Melnikov subharmonic function is then defined as \cite{Guckenheimer1983}:
\begin{equation}
\label{eq:Melnikov}
    M_{m,n}(t_0)=\int_0^{nT_d}\{H_0,H_1\}(q^{E}(t),t+t_0)dt,
\end{equation}
where $q^{E}(t)=(\rho(t),p_\rho(t))$ is a solution of the unperturbed system of energy $E$, fulfilling the resonance condition (\ref{eq:resonance}). If $M_{m,n}(t_0)$ has a simple zero, then for sufficiently small $\epsilon$, there exists a subharmonic periodic orbit of period $n T_d$ of the perturbed system near the corresponding unperturbed periodic orbit. For such a simple zero of $M_{m,n}$ at $t_0=t_0^*$, the surviving orbit is close to the reference orbit $q^E(t)$ evaluated at $t=t_0^*+k T_d$, for $k=0,...,n-1$. This evaluation of $q^E(t)$ selects $n$ stroboscopic points from the already known $q^E$. Note that only for $\epsilon=1$ does the system describe geodesic motion in the Einstein-Rosen SGW. Melnikov methods only describe a first order perturbation in $\epsilon$, and the full system is recovered only at $\epsilon=1$, so quantitatively, the resulting predictions of the exact location of fixed points of the Poincar\'e map might be unreliable. Nonetheless, searching for simple zeros of the subharmonic Melnikov function for our system can be useful in describing qualitatively the observed behavior of the full system.

The minimum value of energy in the unperturbed system is $E_{min}=H_0(\rho_{c},0)\approx1.539$. This corresponds to the minimum period $T_{min}=T(E_{min})\approx4.185$. Together with eq.\ (\ref{eq:resonance}), this leads to the conclusion that subharmonics of period 4 (and lower) cannot exist for this set of parameters, at least for small $\epsilon$. This is a hint as to why no networks of period 4 or lower were observed in the Poincar\'e map discussed in section \ref{sec:layered}. Because of eq.\ (\ref{eq:resonance}), only subharmonics for which $\frac{n}{m}>T_{min}$ are possible. The first such subharmonic of $m=1$ is $(m,n)=(1,5)$. To find orbits of period $T_{min}<T(E)<5$, one can explore subharmonics with $m>1$. The table \ref{tab:subharmonics} lists the subharmonics for $m=1,2,3$, which create co-prime pairs such that $T_{min}<n/m\leq5$.

\begin{table}[]
    \caption{Co-prime pairs $(m,n)$ such that $T_{min}<n/m\leq5$, ordered by ascending $T$. $T_{min}=4.185$.}
    \begin{ruledtabular}
    \begin{tabular}{cccc}
         m&n&E&T=n/m  \\\colrule
         3&13&1.589&$4.\overline{3}$  \\
         2&9&1.648&4.5  \\
         3&14&1.710&$4.\overline{6}$  \\
         1&5&1.841&5  \\
    \end{tabular}
    \end{ruledtabular}
    \label{tab:subharmonics}
\end{table}

\begin{figure}[]
\centering
\includegraphics[width=0.49\textwidth]{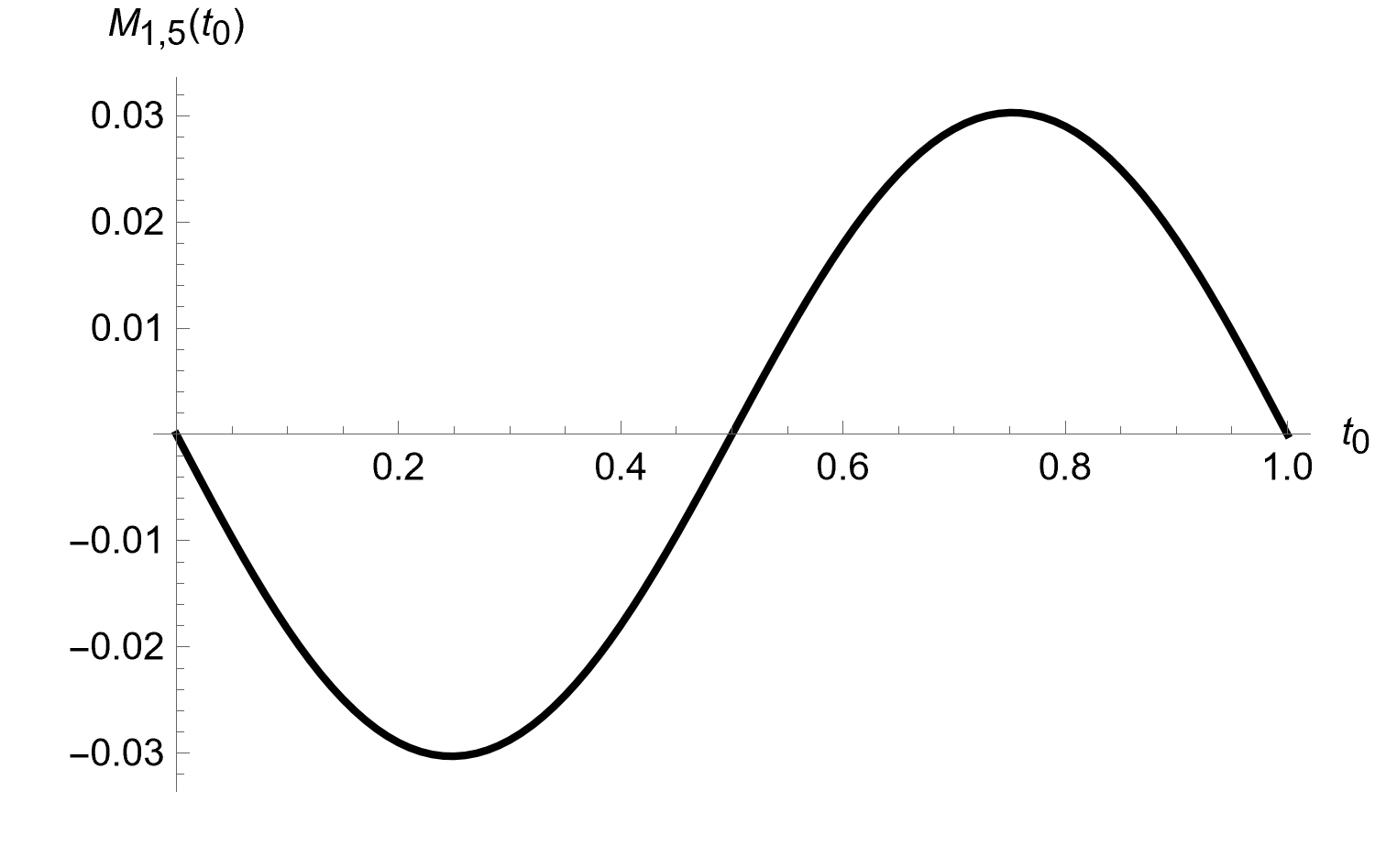}
\includegraphics[width=0.49\textwidth]{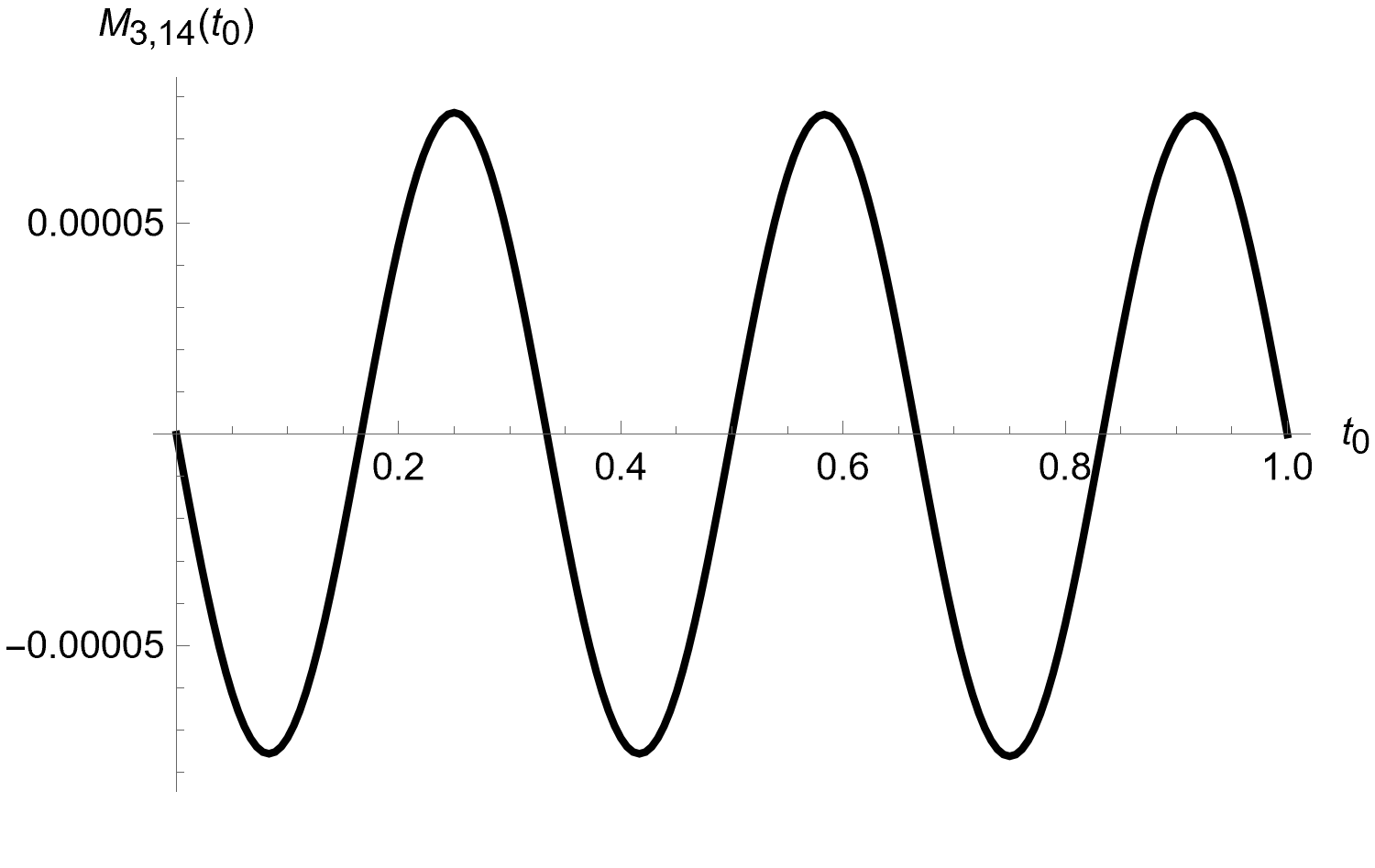}

\caption{Numerically computed $M_{1,5}(t_0)$ (top) and $M_{3,14}(t_0)$ (bottom).}
\label{fig:M15}
\end{figure}

The numerically computed $M_{1,5}$ is presented in Figure~\ref{fig:M15} (top). $M_{1,5}$ has two unique simple zeros at $t_0=0$ and $t_0=0.5$. $M_{m,n}$ is periodic with period $T_d$ by construction, so the zero at $t_0=1$ represents the same subharmonic as $t_0=0$. Each of the two zeros corresponds to a surviving subharmonic trajectory generating a set of five 5-periodic points, which are fixed points of $\phi^5$ (5 hyperbolic and 5 elliptic fixed points in total). This is the subharmonic responsible for the 5-periodic network shown in black in the Poincar\'e map in Figure \ref{fig:layered_external}. A similar analysis for $m=1$ and $n=6,7,8...$ recovers the outer layers of points of periodicity $6,7,8,...$ presented earlier in Figure \ref{fig:layered_external}. The same analysis was done for the candidate subharmonics for $m=2$ and $m=3$ listed in table~\ref{tab:subharmonics}. The results are very similar for different subharmonics, so only $M_{3,14}(t_0)$ was chosen as one more example, and its numerically obtained plot is presented in Figure \ref{fig:M15} (bottom). $M_{3,14}(t_0)$ appears to have more simple zeros, however zeros related by $t_0\rightarrow t_0+\frac{n}{m}T_d\,\text{mod}\,1$ belong to the same closed orbit, so for a given pair $(m,n)$, the $2m$ zeros always describe one elliptic and one hyperbolic resonant chain of $n$ points each. The subharmonics listed in table \ref{tab:subharmonics} are exactly what was observed for the full perturbed system with the help of the Poincar\'e map: inside the 5-periodic heteroclinic network, many more were observed, starting with period 14, 9, and then 13, as presented earlier in Figure \ref{fig:layered_internal}. Despite the Melnikov function being a first order method in $\epsilon$, and the original system corresponding to $\epsilon=1$, the behavior described by subharmonic Melnikov functions is qualitatively consistent with what is being observed in the full system.

Until now, this work has focused on a set of parameters first picked by the authors of \cite{SzybkaNaqvi2023chaos} ($A=1$, $\lambda=\frac{1}{2\pi}$, $l=1/3$). This choice was preserved to facilitate easier comparison and to allow for direct extension of the results presented therein. In an attempt to better understand the chaotic properties of the system, the following section will go outside of this set of parameters and examine the behavior of the system and the Poincar\'e section related to it under a continuous change of parameters.

\section{Evolution of structures in the high-frequency limit} \label{sec:evolution}
\label{sec:HFL}
\begin{figure*}[]
\centering
\includegraphics[width=0.3\textwidth]{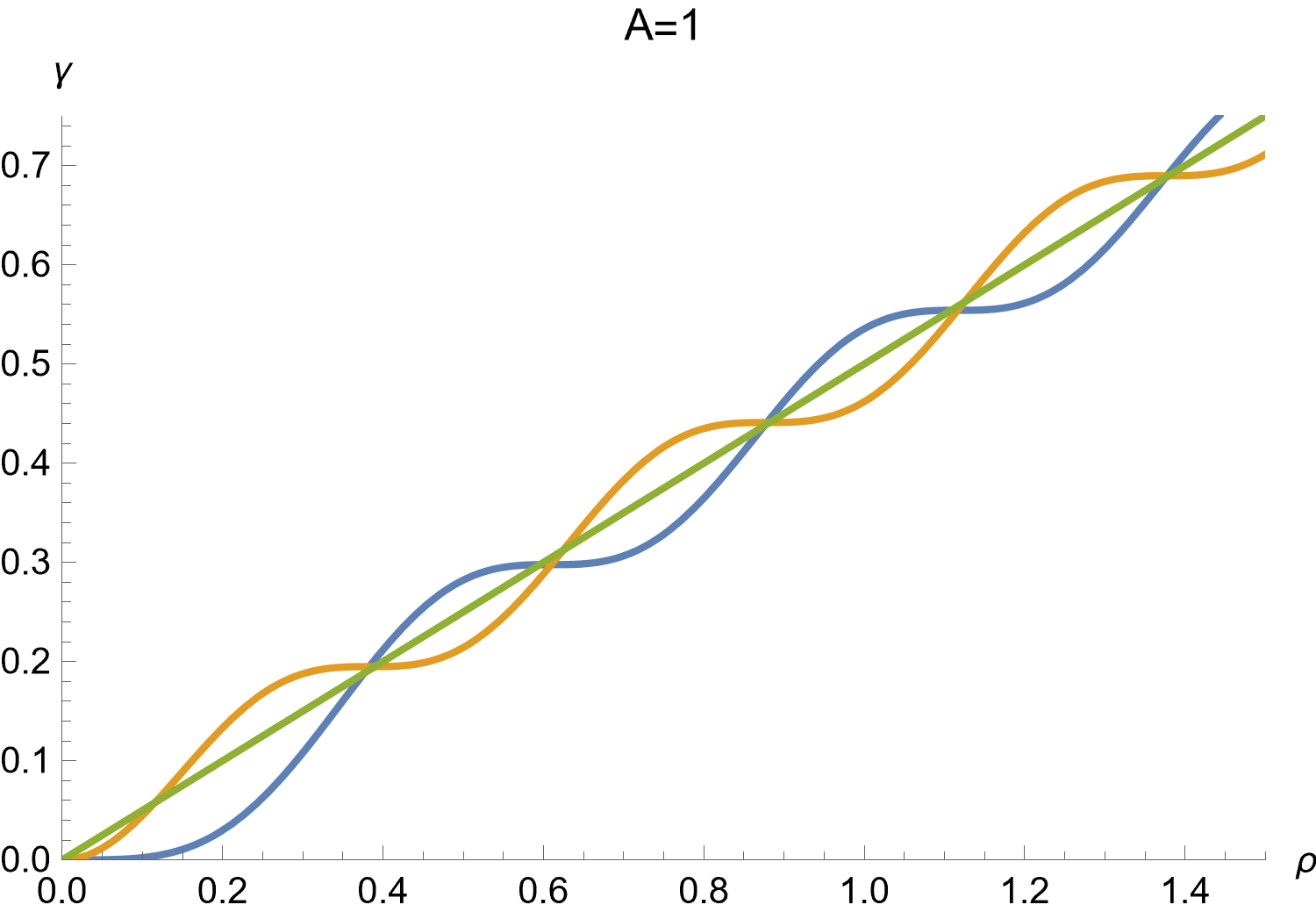}
\includegraphics[width=0.3\textwidth]{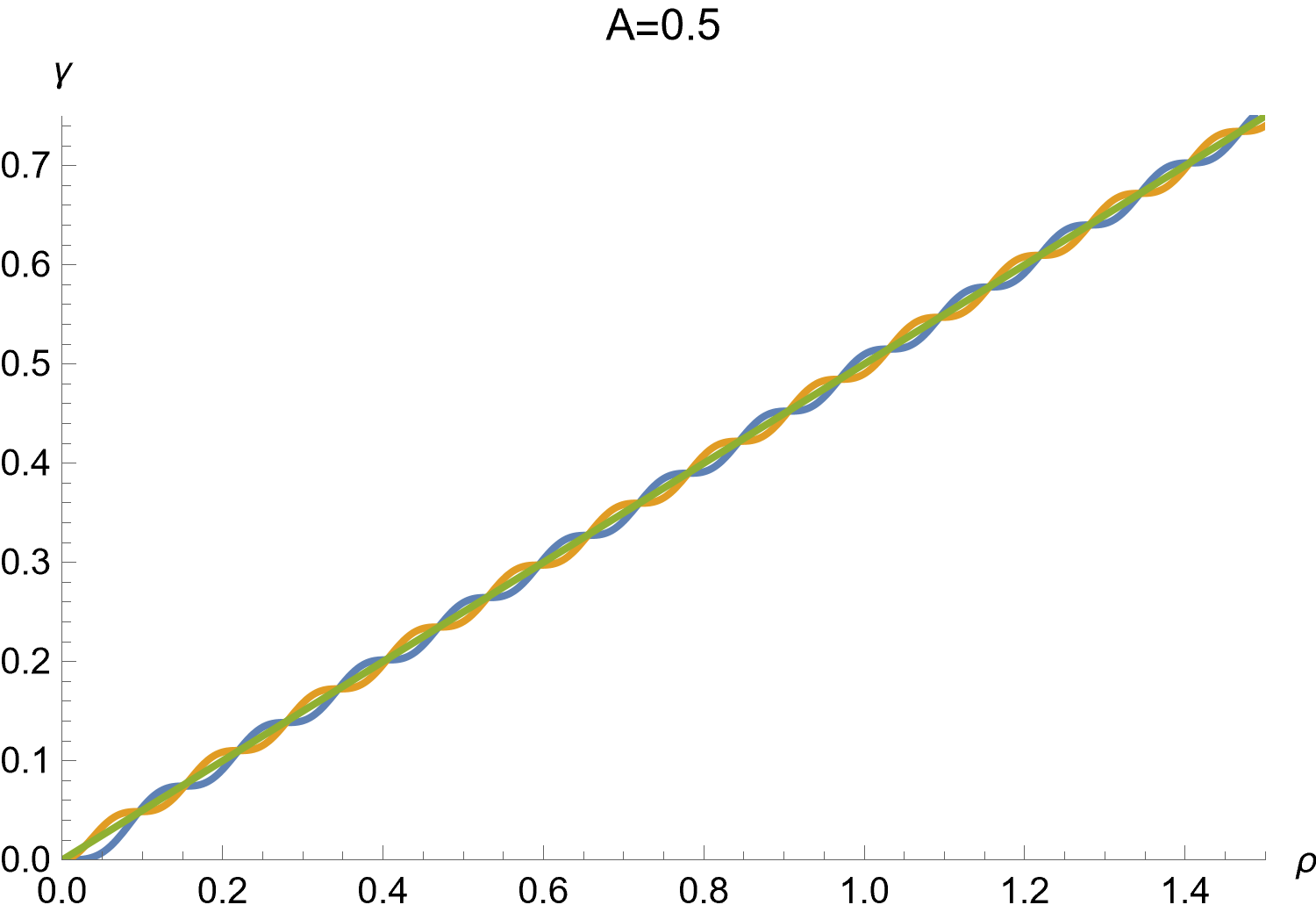}
\includegraphics[width=0.3\textwidth]{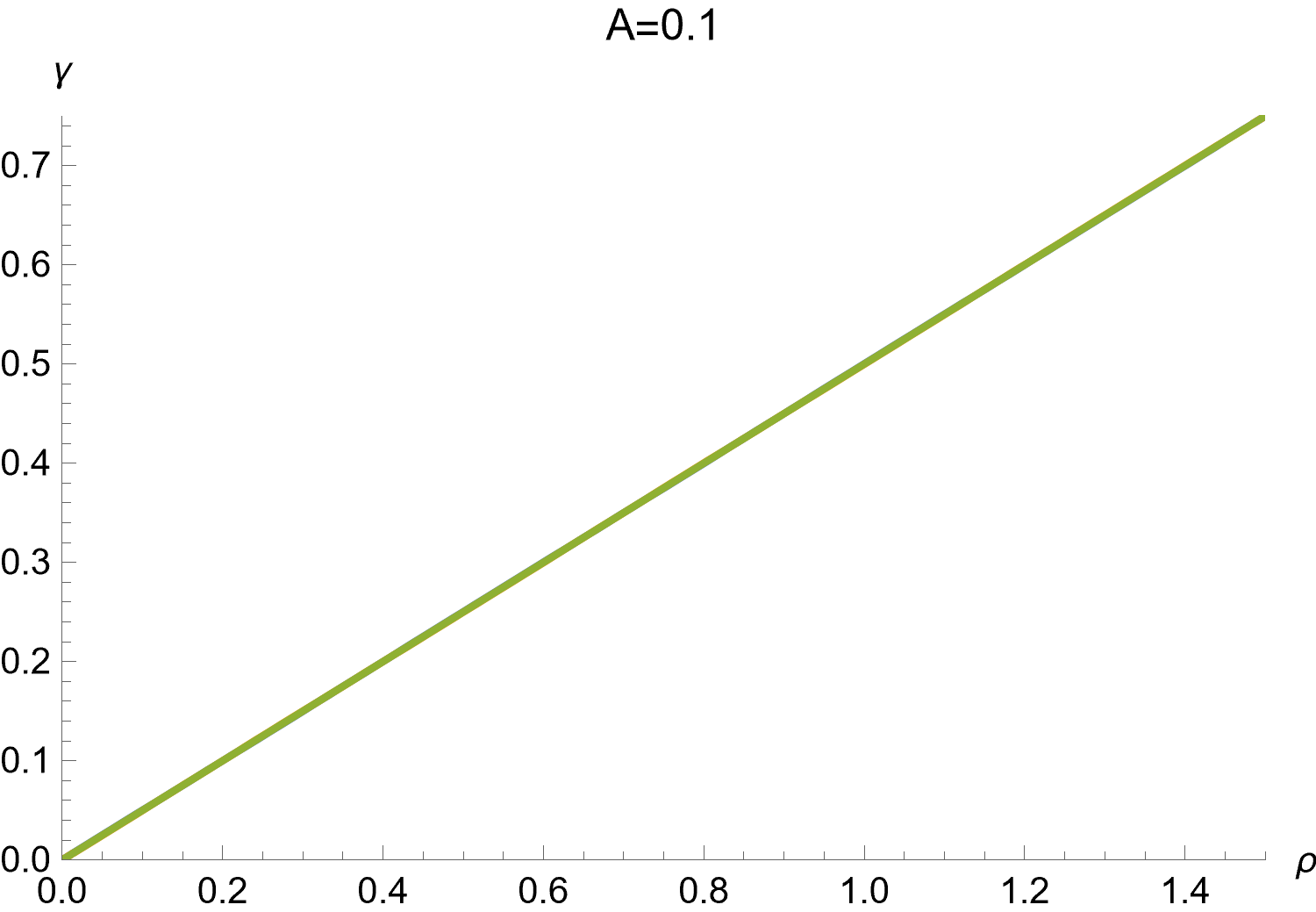}
\caption{Example plots of $\gamma(t,\rho)$ for different values of $A$ and $\lambda=\frac{A^2}{4\beta^2}$. The yellow and blue lines represent the function $\gamma$ at different phases of its periodic oscillation (precisely: at $t/\lambda=0$ and $t/\lambda=\pi/2$). For small values of $A$, the values quickly approach the limiting case (eq.\ \ref{eq:gamma_limit}), shown in green.}
\label{fig:gammaA}
\end{figure*}

In the previous sections, it was highlighted that the chaotic heteroclinic structures detected in Poincar\'e sections are more elaborate than previously noted. To further explore the chaotic behavior of trajectories in this standing wave spacetime, one can vary the parameters of the solution and instead examine a family of solutions and a series of Poincar\'e maps. One physically motivated choice of such a family was presented in \cite{backreaction}. There, the authors construct a family of solutions fulfilling the requirements of the Green-Wald backreaction framework \cite{greenwald}. The family is constructed by relating the amplitude $A$ with the wavelength $2\pi\lambda$ by setting $A=2\beta\sqrt{\lambda}$, that is, $\lambda=\frac{A^2}{4\beta^2}$, where $\beta>0$ is a new constant. This scaling of $A$ together with $\lambda$ allows for the construction of the high-frequency limit $A\rightarrow0$ (or equivalently $\lambda\rightarrow0$) within the Green-Wald framework. Imposing this relation of $A$ and $\lambda$, the limit $A\rightarrow0$ leads to:
\begin{equation}
\label{eq:gamma_limit}
    \lim_{A\rightarrow0}\gamma(t,\rho)=\frac{\beta^2 \rho}{\pi},
\end{equation}
\begin{equation}
    \lim_{A\rightarrow0}\psi(t,\rho)=0
\end{equation}
everywhere outside of the symmetry axis. Figure \ref{fig:gammaA} presents plots of $\gamma(t,\rho)$ for different values of $A$ and $\lambda$ belonging to this family of solutions. This limit form given by eq.\ (\ref{eq:gamma_limit}) also closely coincides with the time-average of $\gamma$, providing a good reference point of the ``average C-energy'' at a given radial distance $\rho$ for a given $\beta$. I picked $\beta^2=\pi/2$, so that $A=1$ corresponds to $\lambda=(2\pi)^{-1}$, which coincides with the parameters used in all previous sections. The metric resulting from this kind of high-frequency limit of the metric (\ref{eq:metricER}) takes a very simple form:
\begin{equation}
\label{eq:metricHFL}
    ds^2=e^{\frac{2\beta^2\rho}{\pi}}(-dt^2+d\rho^2)+\rho^2d\phi^2+dz^2.
\end{equation}

\begin{figure*}[]
\centering
\includegraphics[width=0.3\textwidth]{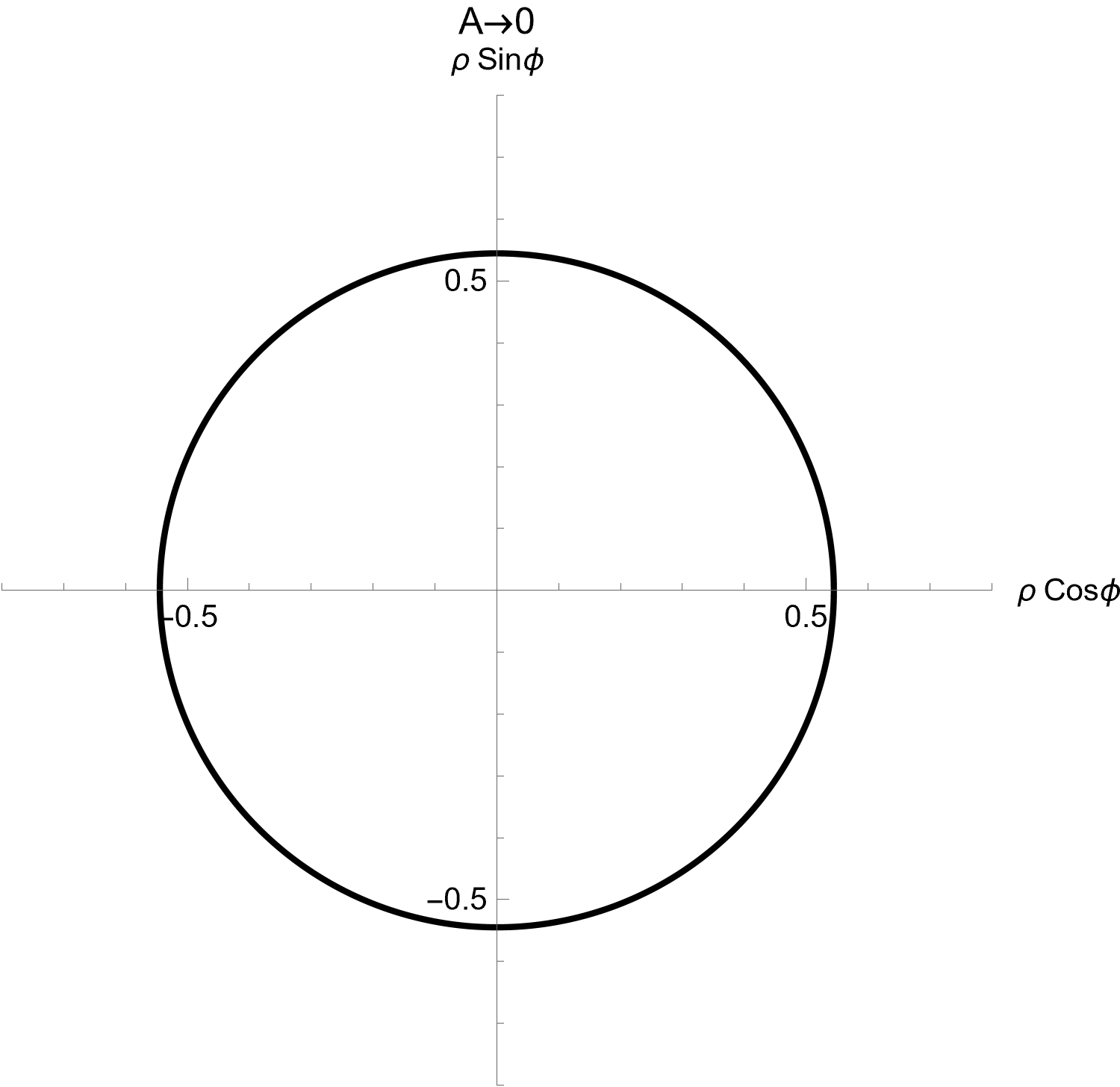}
\includegraphics[width=0.3\textwidth]{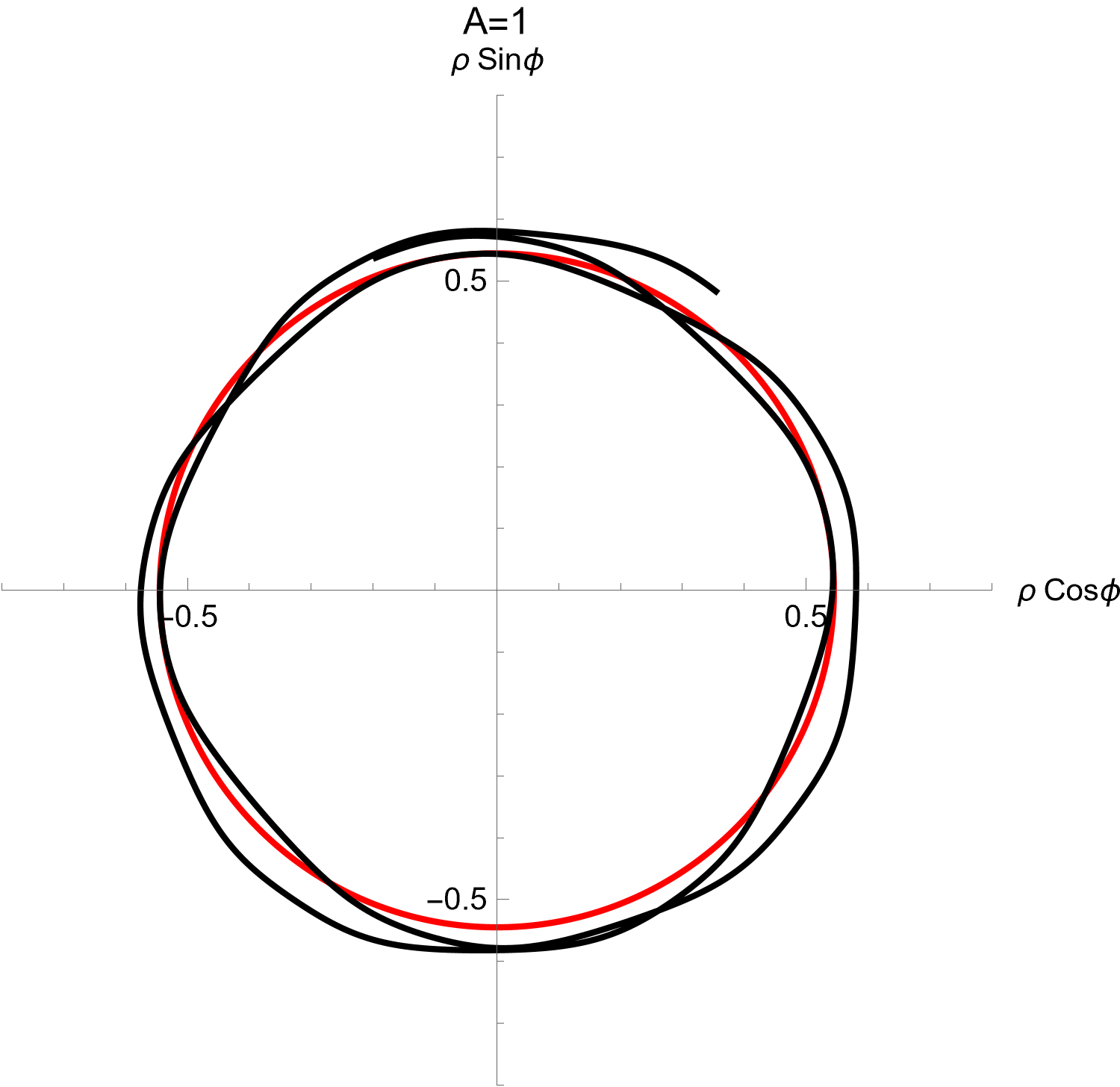}
\includegraphics[width=0.3\textwidth]{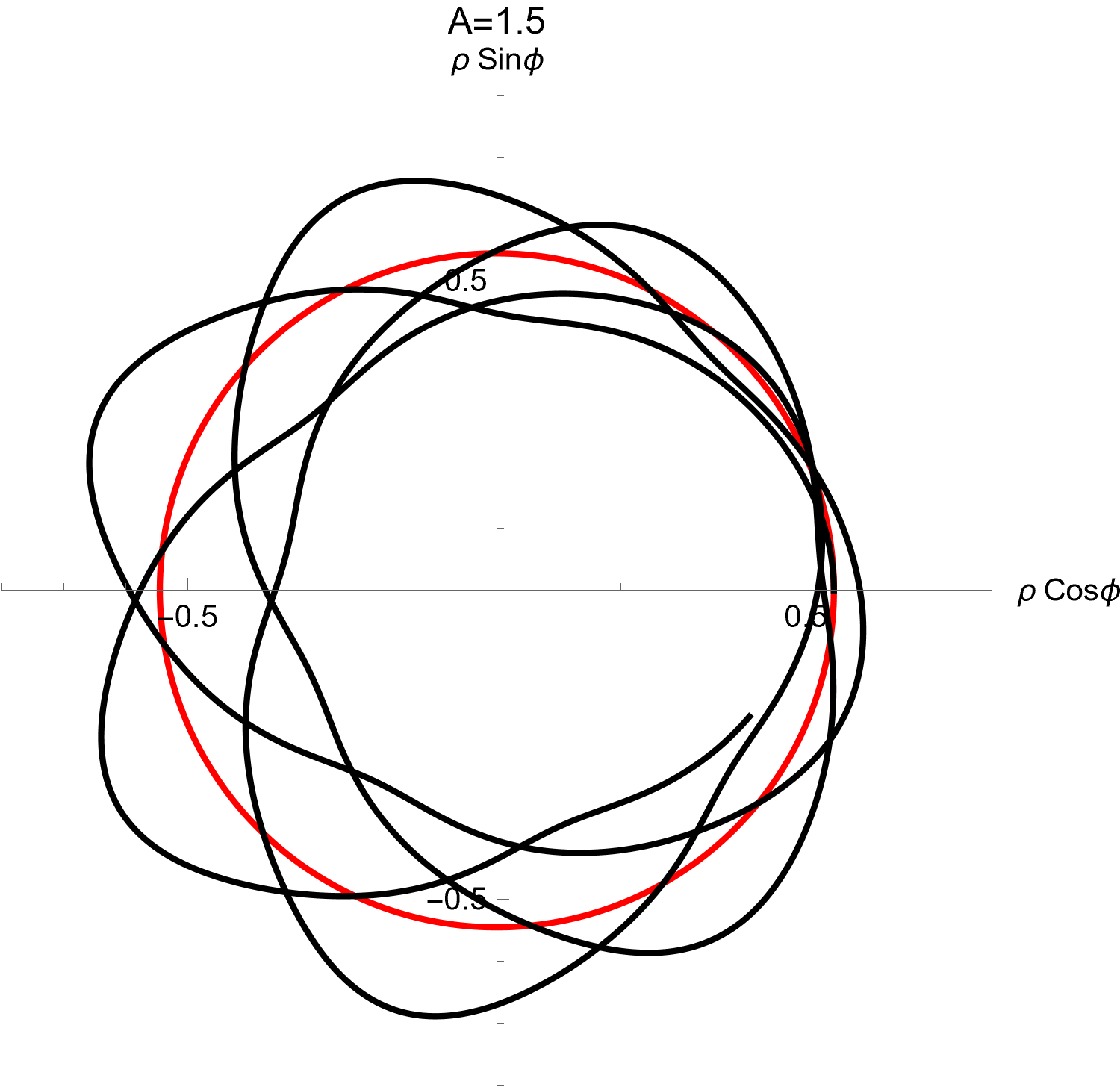}
\caption{Example trajectories of a test particle orbiting the gravitating center for different values of the amplitude $A$ (0, 1, 1.5). In the limiting case, for a given angular momentum $l$, a closed circular orbit can be found (shown in red, also corresponding to the red point in Figures \ref{fig:Veff} and \ref{fig:phasespace}). For values of $A>0$, the behavior resembles perturbations of a circular orbit by the periodic radial driving force of the gravitational wave.}
\label{fig:perturb}
\end{figure*}

The metric (\ref{eq:metricHFL}) is no longer a vacuum solution and does not belong to the Einstein-Rosen class itself. Instead, as discussed in \cite{backreaction}, this spacetime can be thought of as filled with null fluid. This work focuses on solutions with \mbox{$A>0$}, which are vacuum solutions for each value of $A$. With this choice of the family of solutions, the behavior of test particles resembles motion under the effects of a periodic radial force perturbing the movement of test particles about a cylindrically symmetric gravitating center. The larger the value of $A$, the stronger the perturbation, as demonstrated in Figure \ref{fig:perturb}.

\begin{figure*}[]
\centering
    \begin{center}
    \frame{\includegraphics[width=0.44\textwidth]{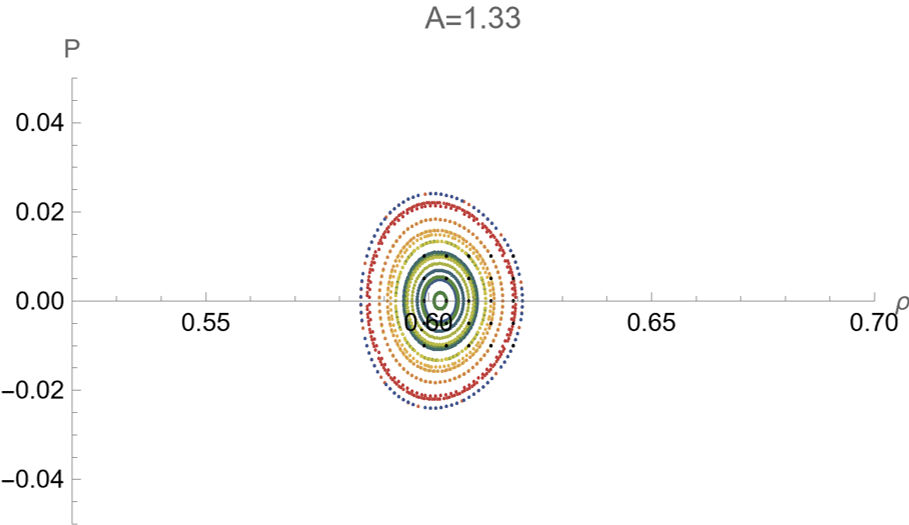}}
    \frame{\includegraphics[width=0.44\textwidth]{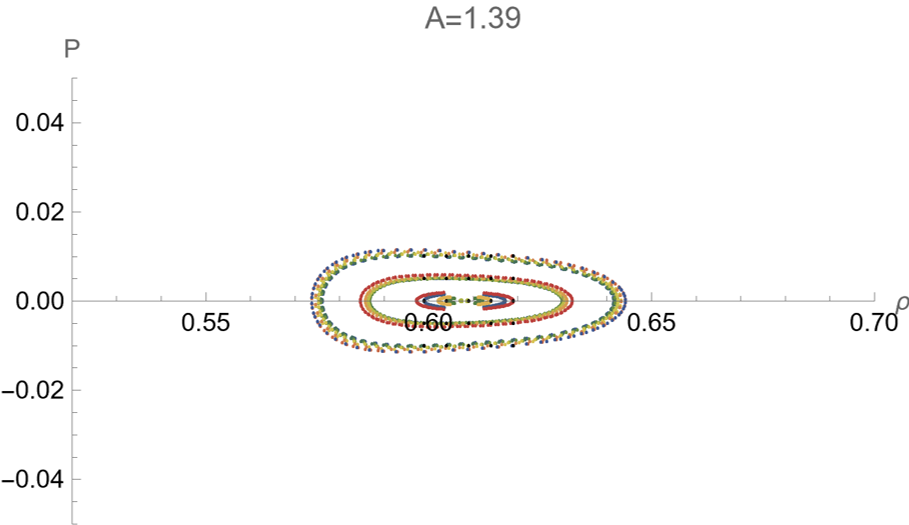}}
    \end{center}
    \vspace{-0.57cm}
    \begin{center}
    \frame{\includegraphics[width=0.44\textwidth]{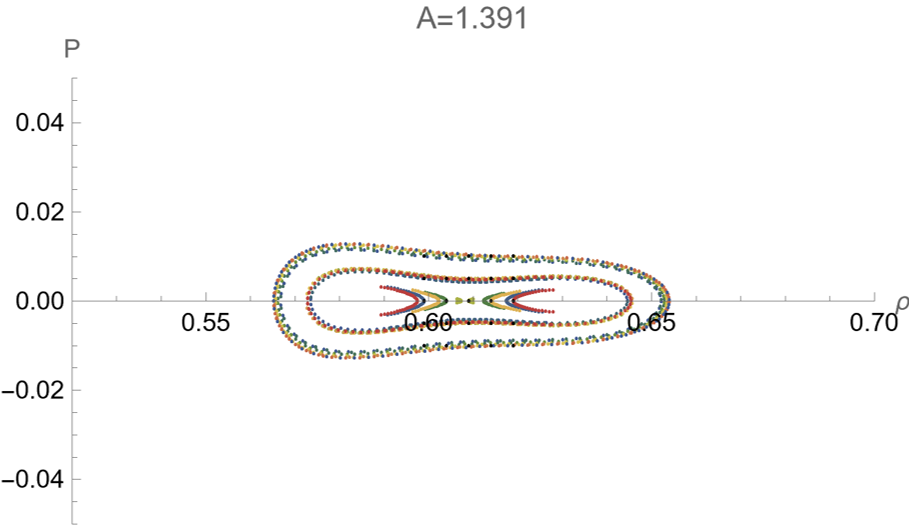}}
    \frame{\includegraphics[width=0.44\textwidth]{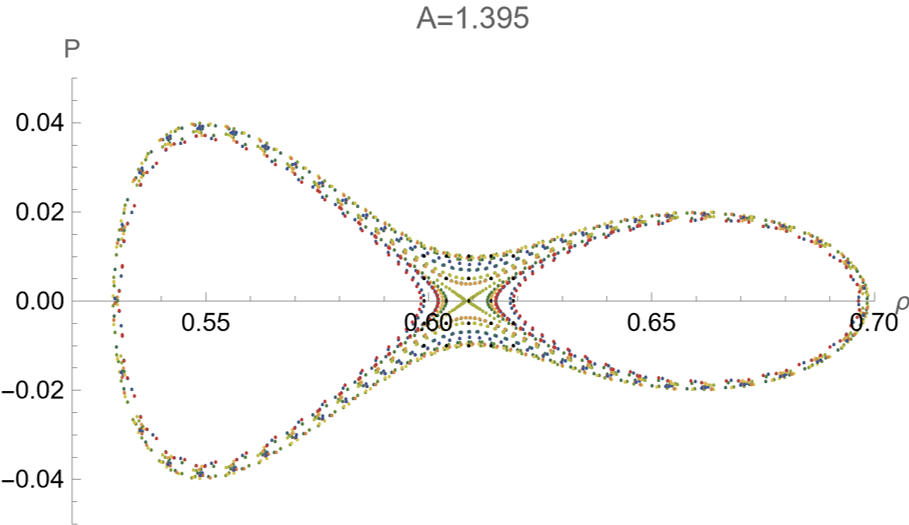}}
    \end{center}
\caption{The period-doubling bifurcation of the fixed point $p_c$ at $A\approx1.3905$. Different colors denote different initial conditions.}
\label{fig:BFP}
\end{figure*}

\begin{figure}[]
\centering
\includegraphics[width=0.4\textwidth]{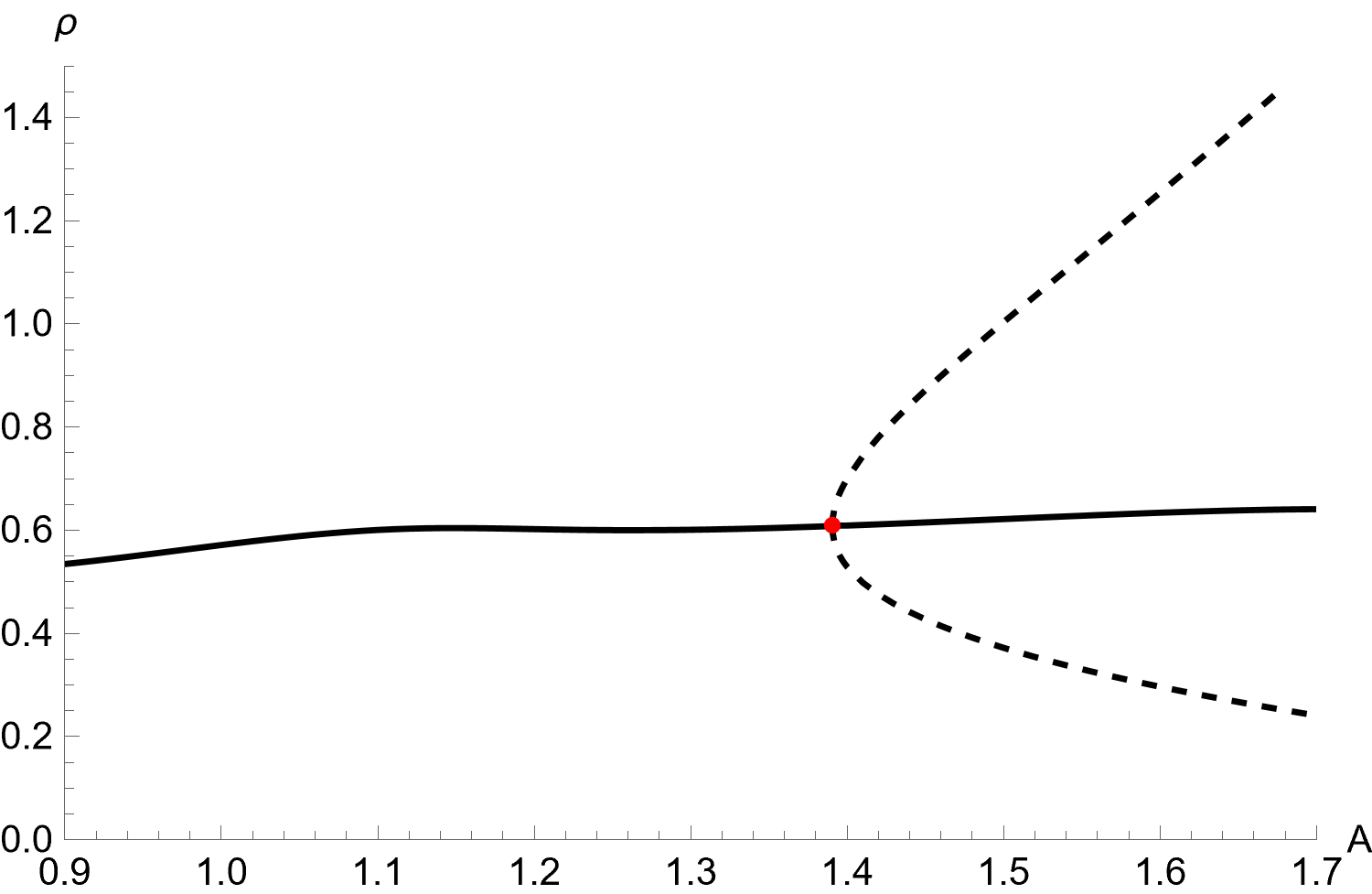}

\caption{Numerically found positions of fixed points along the axis $P=0$ as a function of the amplitude parameter $A$. The period-doubling bifurcation is occurring at $A\approx1.3905$, marked by a red dot. Solid line - the position of fixed point $p_c$, dashed line - the position of 2-periodic points.}
\label{fig:bifurcation}
\end{figure}

Using a scaling motivated by the high-frequency limit and a constant time-average of the C-energy, the effects of changing the value of $A$ on the Poincar\'e map were examined. Different values of the $A$ parameter, now related to both the amplitude and wavelength of the wave, lead to a similar macroscopic picture of the Poincar\'e map. Figures \ref{fig:PS_part1} and \ref{fig:PS_part2} show the obtained Poincar\'e sections for various values of the parameter $A$. As visible in Figures \ref{fig:PS_part1} and \ref{fig:PS_part2}, the general location of the elliptic fixed point $p_c$ is preserved, with slight displacements. The scale of the features appears to be similar, but the specifics of the periodic points and the heteroclinic networks they belong to vary. Different periodic points arise and disappear with a continuous change of the parameter $A$. I~picked one such occurrence for closer examination. Between the values of $A=1.3$ and $A=1.4$, the central fixed point $p_c$ appears to be disrupted. A precise examination shows a period-doubling bifurcation of the fixed point $p_c$. It changes from elliptic to hyperbolic, developing two homoclinic loops, while a new period-2 orbit is born. The two new 2-periodic points are elliptic fixed points of the map $\phi^2$. The bifurcation of the fixed point $p_c$ is shown in Figure \ref{fig:BFP}. The evolution of the position and number of fixed points with a continuous change in the parameter $A$ is presented in Figure \ref{fig:bifurcation}.

\section{Summary}
I explored the chaotic behavior of geodesics in an Einstein-Rosen standing gravitational wave spacetime. I extended the results of \cite{SzybkaNaqvi2023chaos} by demonstrating that the underlying structures detectable in Poincar\'e maps are much richer than previously reported. Multiple heteroclinic networks of similar behavior, but different sizes and periodicity were identified. In this way, I established that standing gravitational waves can produce complex dynamics of test particle motion even in the case of one of the simplest, cylindrically symmetric SGW solutions.

The conducted analysis demonstrates how the subharmonic Melnikov function method can be a useful tool for obtaining a better understanding of the dynamics compared to relying only on direct examination of Poincar\'e sections. Using the subharmonic Melnikov function method, I explained why only heteroclinic networks of specific periodicity appear in the Poincar\'e map.

Putting the studied SGW solution in the context of the results obtained in \cite{backreaction}, I explored the change of test particle dynamics along the high-frequency limit motivated by the Green-Wald backreaction framework. I demonstrated how the resulting Poincar\'e sections evolve with a continuous change of parameters, leading to a period-doubling bifurcation.

\FloatBarrier
\bibliographystyle{apsrev4-2}
\setcitestyle{authortitle}
\bibliography{bib}

\appendix
\section{Additional figures}
\begin{figure*}[t]
\centering
\includegraphics[width=0.49\textwidth]{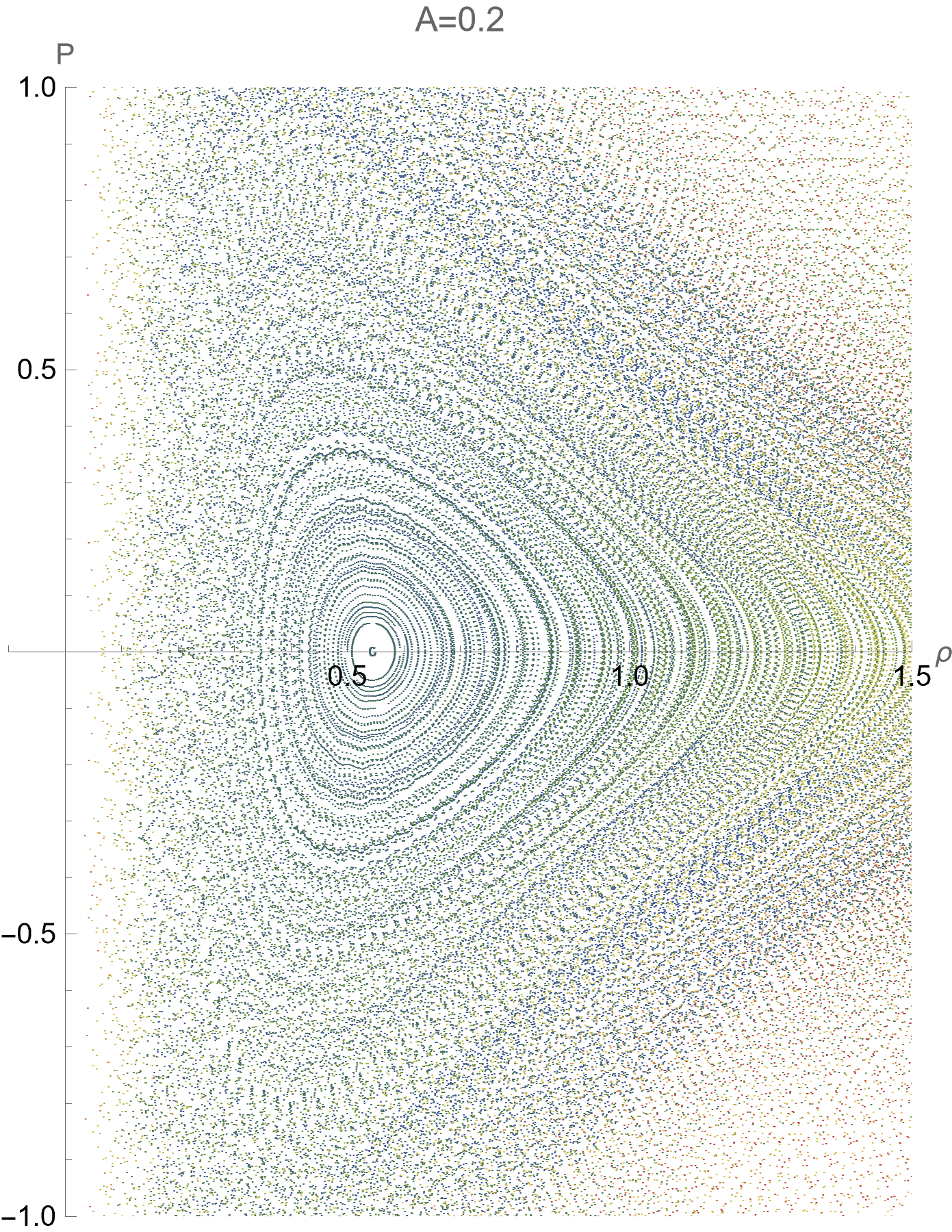}
\includegraphics[width=0.49\textwidth]{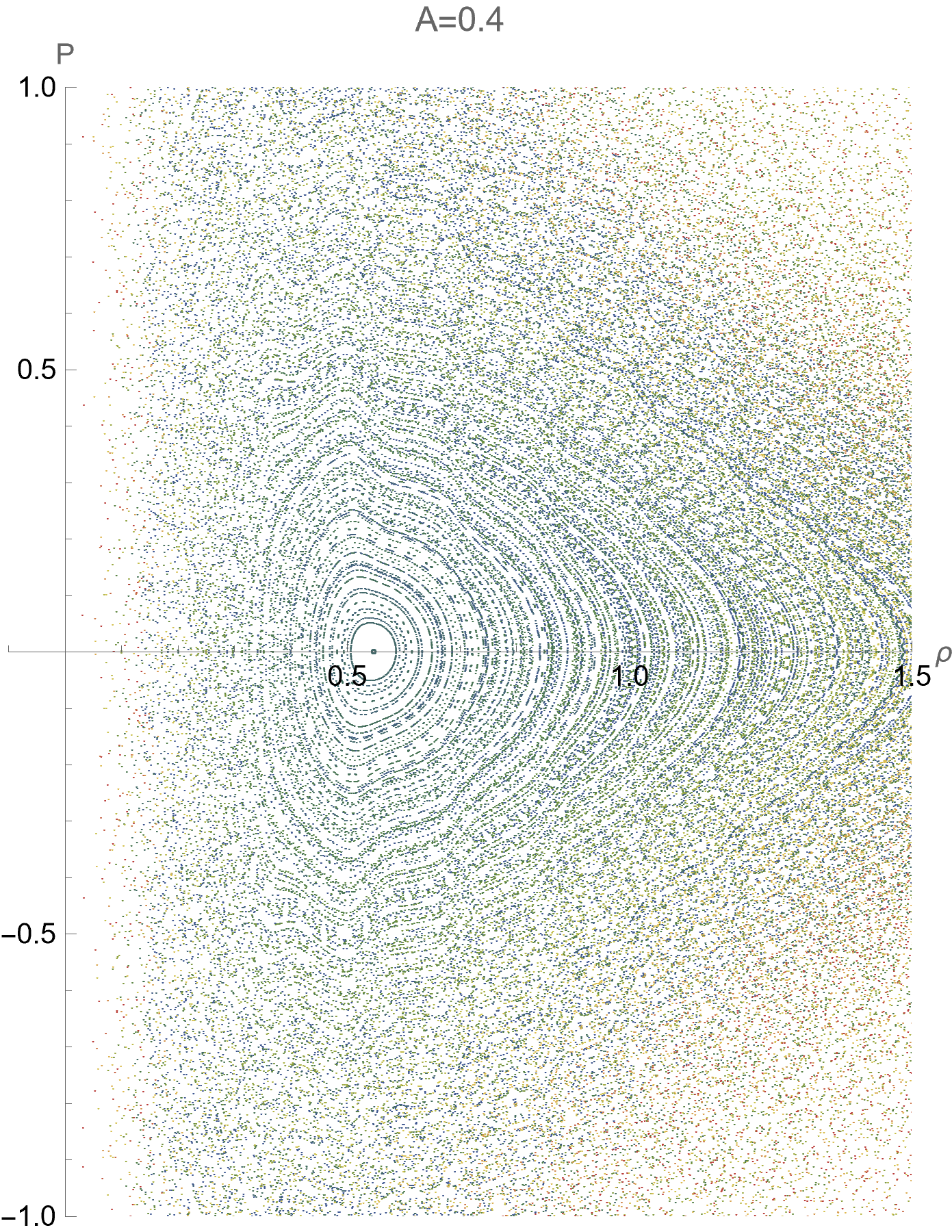}\\[0.5em]
\includegraphics[width=0.49\textwidth]{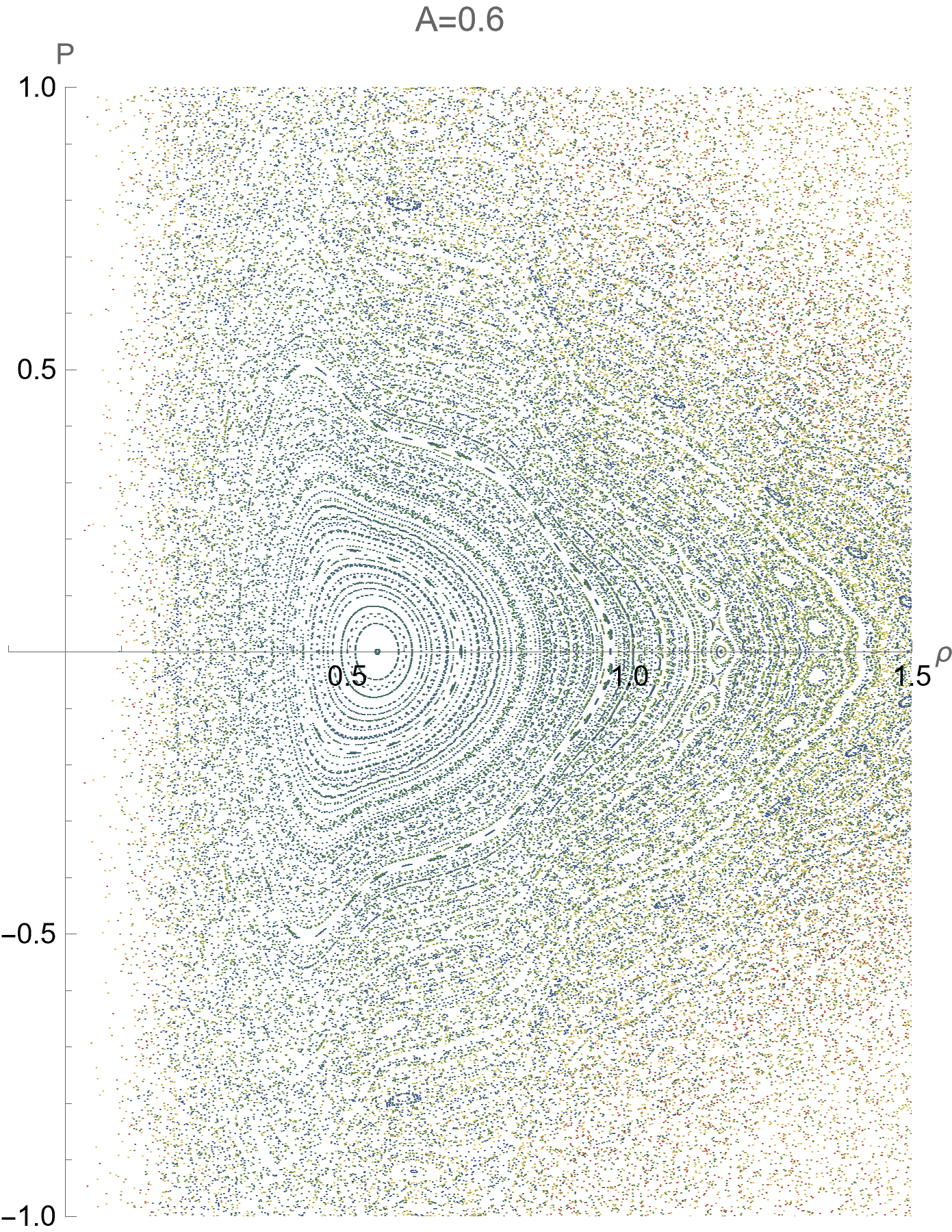}
\includegraphics[width=0.49\textwidth]{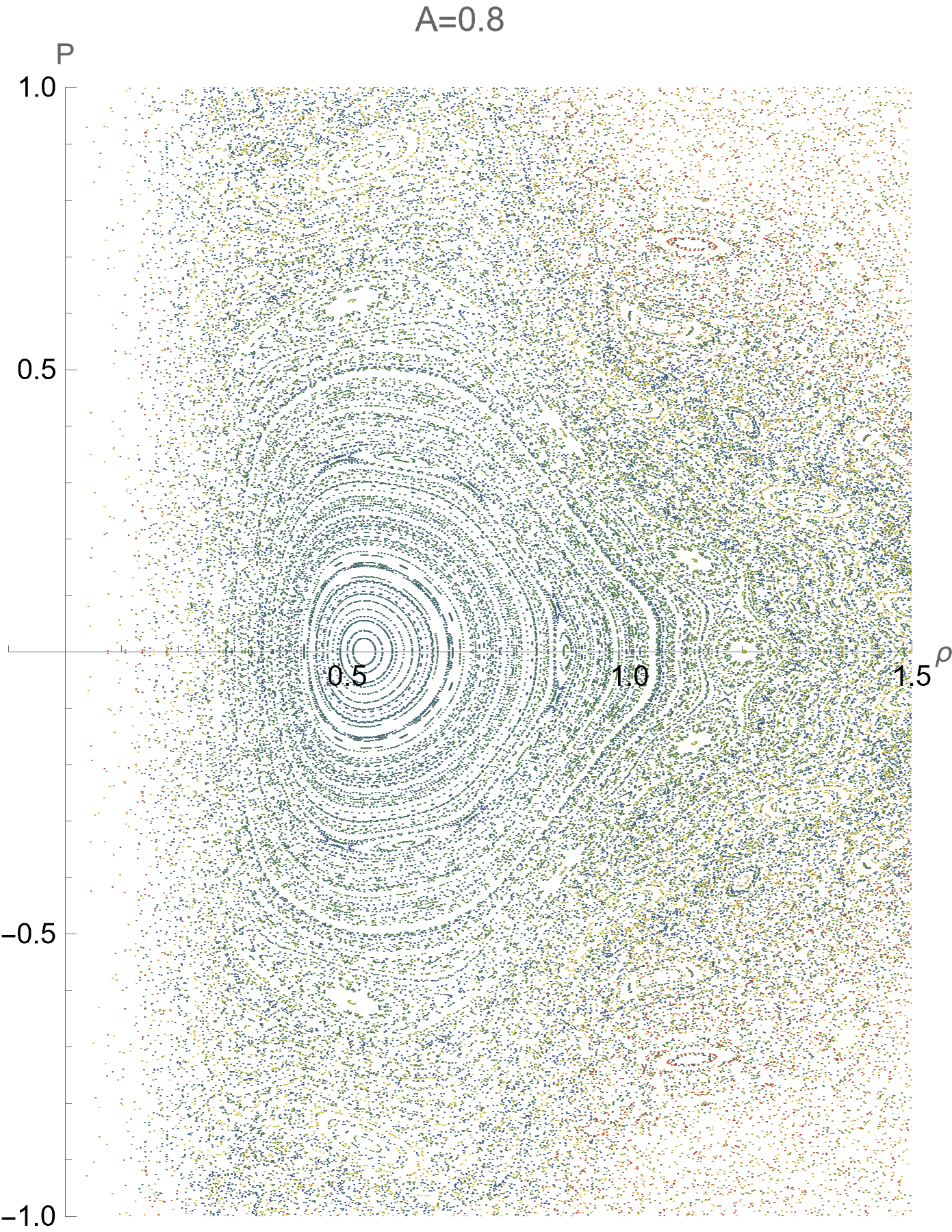}
\caption{The Poincar\'e sections of $t = 2\pi \lambda k$, for $k=0,1,2,...$, for different values of $A$.}
\label{fig:PS_part1}
\end{figure*}

\begin{figure*}[]
\centering
\includegraphics[width=0.49\textwidth]{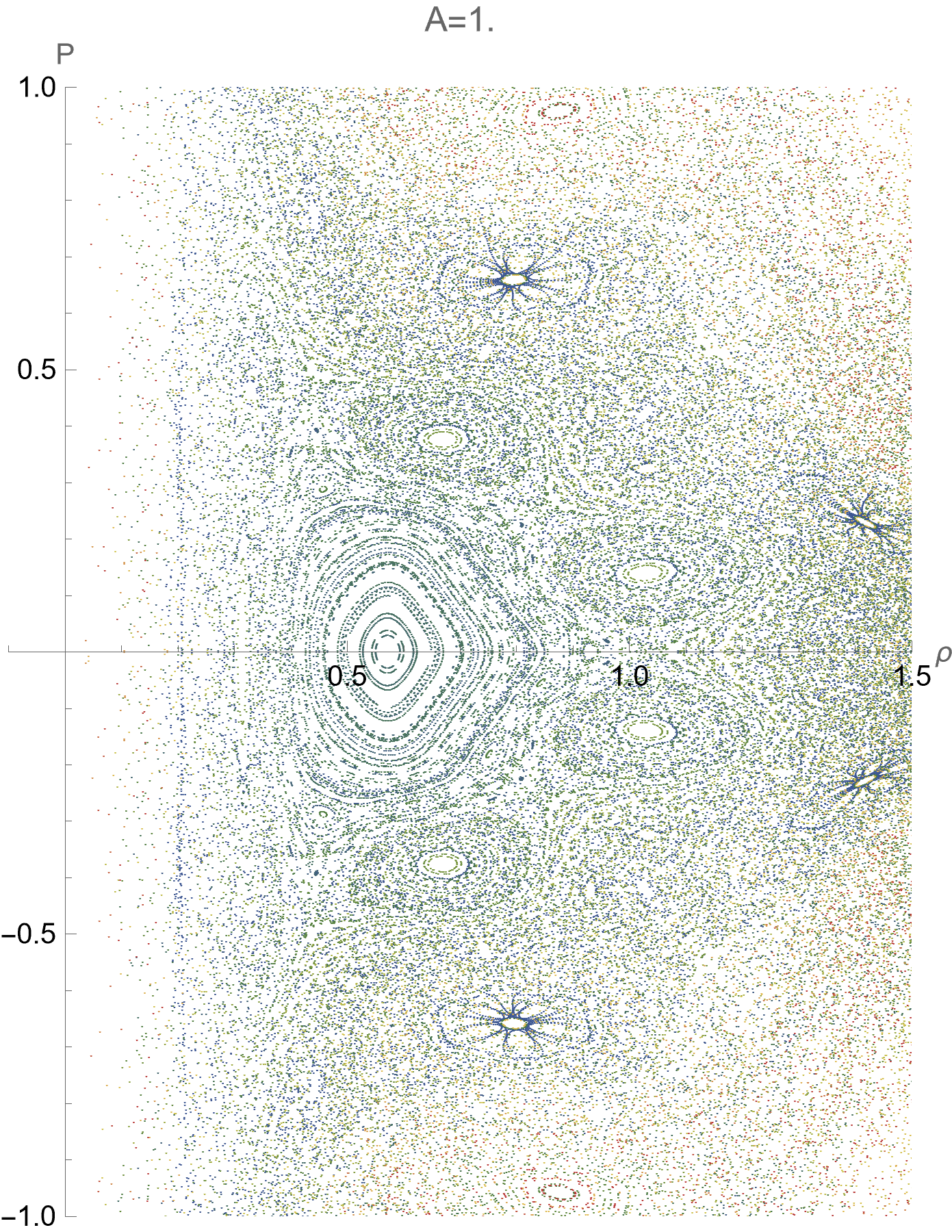}
\includegraphics[width=0.49\textwidth]{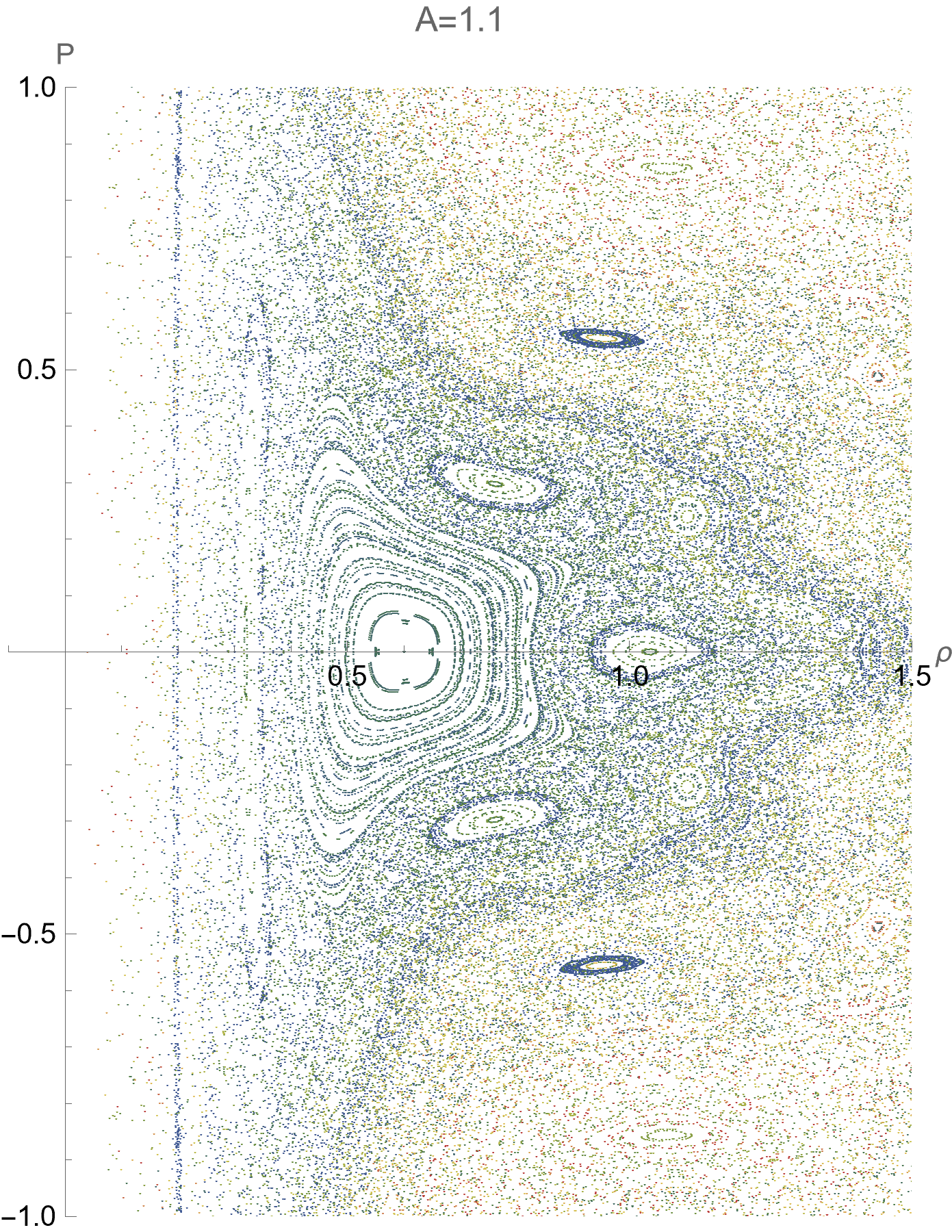}\\[0.5em]
\includegraphics[width=0.49\textwidth]{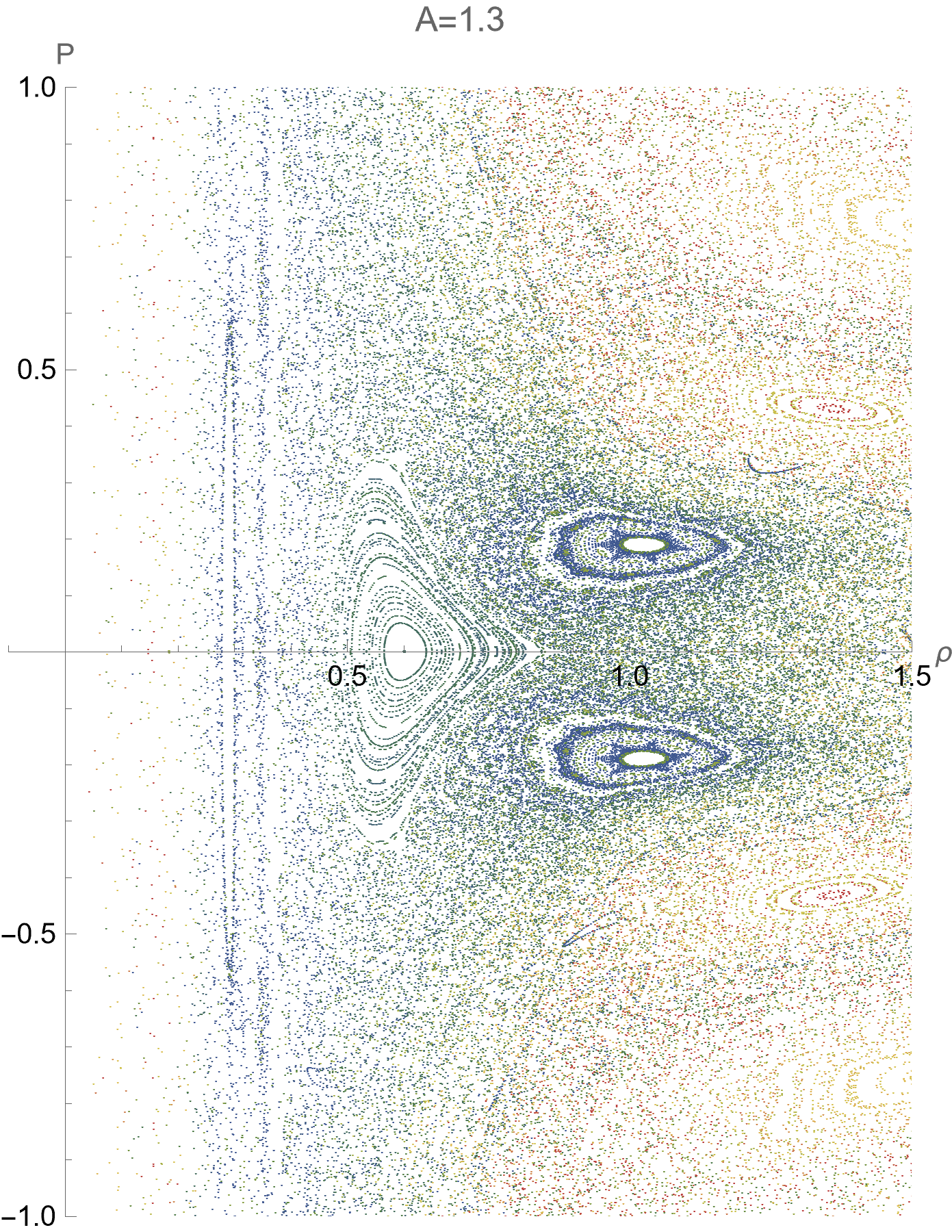}
\includegraphics[width=0.49\textwidth]{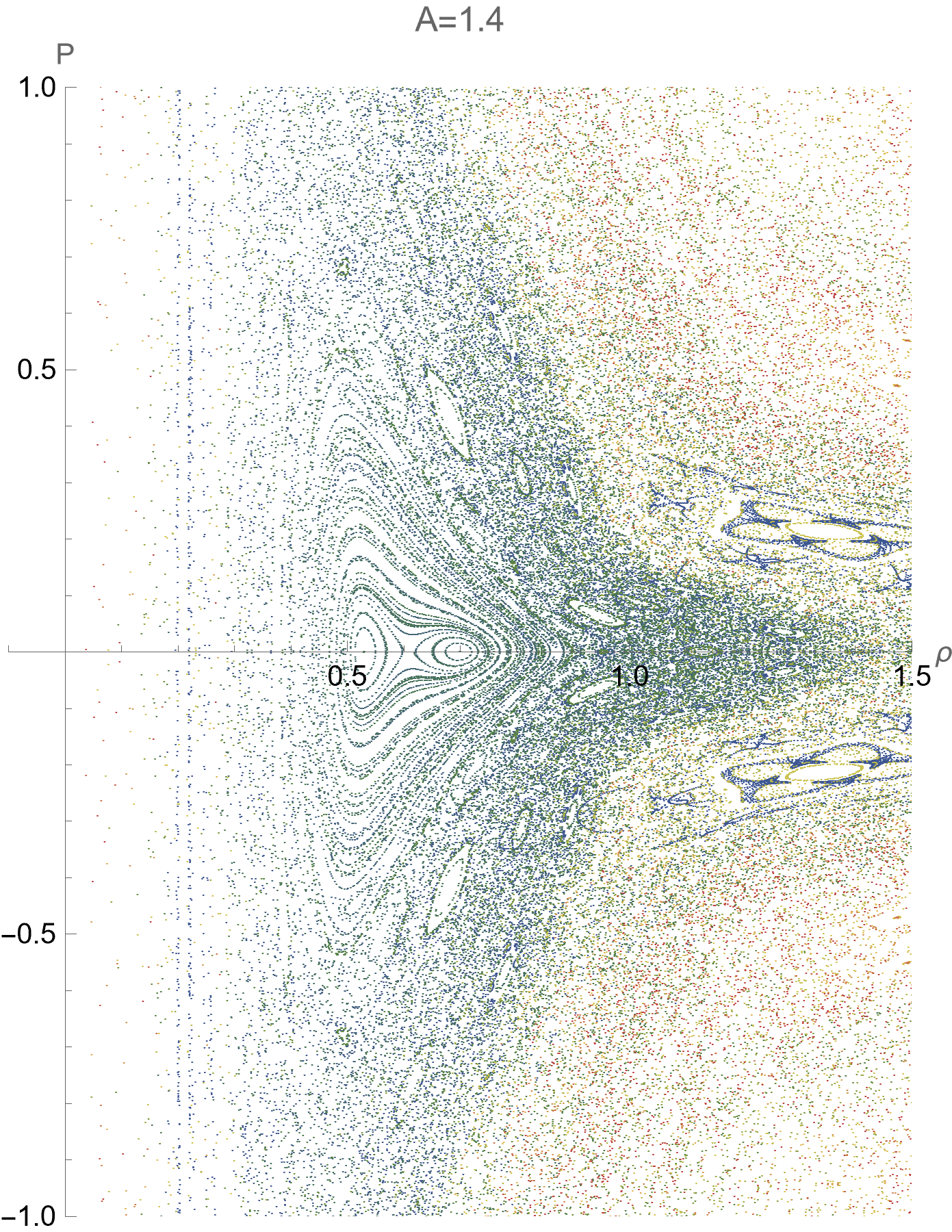}
\caption{The Poincar\'e sections of $t = 2\pi \lambda k$, for $k=0,1,2,...$, for different values of $A$.}
\label{fig:PS_part2}
\end{figure*}

\end{document}